\documentclass[acmsmall,screen]{acmart}
\pdfoutput=1
\usepackage{mathtools}

\AtBeginDocument{%
  }

\usepackage[frozencache]{minted2}
\usepackage{prettyref}
\newrefformat{sec}{Section~\ref{#1}}
\newrefformat{alg}{Algorithm~\ref{#1}}
\newrefformat{fig}{Figure~\ref{#1}}
\newrefformat{tab}{Table~\ref{#1}}
\newrefformat{def}{Definition~\ref{#1}}
\newrefformat{app}{Appendix~\ref{#1}}
\newrefformat{cor}{Corollary~\ref{#1}}
\newrefformat{li}{Line~\ref{#1}}
\newcommand\pref[1]{\prettyref{#1}}
\newcommand\abs[1]{\vert #1 \vert}

\usepackage{listings}
\lstdefinestyle{myLuastyle}
{
    language         = {[5.0]Lua},
    basicstyle       = \ttfamily,
    showstringspaces = false,
    upquote          = true,
}
\usepackage[noend,ruled,linesnumbered]{algorithm2e}
\usepackage{subcaption}
\usepackage{tikz}
\tikzset{ gnode/.style={draw,circle}}
\tikzset{gcnode/.style={draw,rectangle}}

\usepackage{pgf}
\usepackage{pgfplots}

\newcommand\uto{\leftrightarrow}

\newcommand\ETSingleton{\mathtt{ETSingleton}}
\newcommand\ETCut{\mathtt{ETCut}}
\newcommand\ETJoin{\mathtt{ETLink}}
\newcommand\ETLink{\mathtt{ETLink}}

\newcommand\ETNext{\mathtt{ETNext}}
\newcommand\ETPath{\mathtt{ETPath}}
\newcommand\ETParent{\mathtt{ETParent}}

\newcommand\GCInit{\mathtt{GCInit}}
\newcommand\GCRoot{\mathtt{root}}
\newcommand\GCAllocate{\mathtt{GCAllocate}}
\newcommand\GCInsert{\mathtt{GCInsert}}
\newcommand\GCDelete{\mathtt{GCDelete}}
\newcommand\GCStep{\mathtt{GCStep}}
\newcommand\GCFreeList{\mathtt{GCFreeList}}

\newcommand\DCInsert{\mathtt{DRInsert}}
\newcommand\DCDelete{\mathtt{DRDelete}}
\newcommand\DCConnected{\mathtt{DRReaches}}
\newcommand\DCCreate{\mathtt{DRCreate}}

\newcommand\UDCInit{\mathtt{DCInit}}
\newcommand\UDCInsert{\mathtt{DCInsert}}
\newcommand\UDCDelete{\mathtt{DCDelete}}
\newcommand\UDCConnected{\mathtt{DCReaches}}

\newcommand\LPCInsert{\mathtt{LPRInsert}}
\newcommand\LPCDelete{\mathtt{LPRDelete}}
\newcommand\LPCConnected{\mathtt{LPRReaches}}
\newcommand\LPCCreate{\mathtt{LPRCreate}}

\renewcommand\O[1]{O\left(#1\right)}
\newcommand\TOmega{\tilde{\Omega}}

\newcommand\NMax{n_{\mathrm{max}}}

\newcommand\tAR{t_{\mathrm{AR}}}
\newcommand\tAAR{t_{\mathrm{A,AR}}}
\newcommand\tSAAR{t_{\mathrm{S,A,AR}}}
\newcommand\tkSAAR{t_{\mathrm{mS,A,AR}}}

\setminted{fontsize=\small}


\begin{document}

\title[Pathological Cases for a Class of Reachability-Based Garbage Collectors]{Pathological Cases for a Class of Reachability-Based \\ Garbage Collectors}

\author{Matthew Sotoudeh}
\orcid{0000-0003-2060-1009}
\affiliation{%
  \institution{Stanford University}
  \city{Stanford}
  \country{USA}
}
\email{sotoudeh@stanford.edu}


\begin{abstract}
    Although existing garbage collectors (GCs) perform extremely well on
    typical programs, there still exist pathological programs for which modern
    GCs significantly degrade performance.
    This observation begs the question: might there exist a `holy grail' GC
    algorithm, as yet undiscovered, guaranteeing both constant-length pause
    times and that memory is collected promptly upon becoming unreachable?
    For decades, researchers have understood that such a GC is \emph{not}
    always possible, i.e., some pathological behavior is unavoidable when the
    program can make heap cycles and operates near the memory limit, regardless
    of the GC algorithm used.
    However, this understanding has until now been only informal, lacking a
    rigorous formal proof.

    This paper complements that informal understanding with a rigorous
    proof, showing with mathematical certainty that \emph{every} GC algorithm
    that can implement a realistic mutator-observer interface has some
    pathological program that forces it to either introduce a long GC pause
    into program execution or reject an allocation even though there is
    available space.
    Hence, language designers must either accept these pathological scenarios
    and design heuristic approaches that minimize their impact (e.g.,
    generational collection), or restrict programs and environments to a strict
    subset of the behaviors allowed by our mutator-observer--style interface
    (e.g., by enforcing a type system that disallows cycles or overprovisioning
    memory).

    We do not expect this paper to have any effect on garbage collection
    practice.
    Instead, it provides the first mathematically rigorous answers to these
    interesting questions about the limits of garbage collection.
    We do so via rigorous reductions between GC and the dynamic graph
    reachability problem in complexity theory, so future algorithms and lower
    bounds from either community transfer to the other via our reductions.

    We end by describing how to adapt techniques from the graph data structures
    community to build a garbage collector making worst-case guarantees that
    improve performance on our motivating, pathologically memory-constrained
    scenarios, but in practice find too much overhead to recommend for typical
    use.
\end{abstract}



\maketitle

\section{Introduction}
Managing limited memory resources is a fundamental problem in programming
language design.
Garbage collection (GC) is a user-friendly approach where the language runtime
automatically deallocates memory regions once they are no longer reachable from
the local and global variables~\citep{handbook}.
Concerns about predictability and pause times have led many low-level
languages to adopt manual memory management, where the programmer explicitly
tells the runtime when to release resources~\citep{ziggc}.
Other languages use type systems that allow the compiler to predict statically
where to release memory resources at the cost of restricting program
expressiveness~\citep{rustgc}.
This paper explains in a formal way why these tradeoffs are necessary, by proving hard
limits on the asymptotic worst-case performance of a wide class of GCs.

We first give a formal model characterizing precisely the class of garbage
collectors our results apply to.
In particular, we address only those GC algorithms that can be used to
implement a mutator-observer interface where the GC algorithm reads a stream of
pointer updates from the program and then must add regions to a free list once
they become unreachable.
In such a setting, one key issue is how long it takes for the GC to add a
region to the free list after it becomes unreachable: we call this the \emph{delay}.
Many real-time settings, e.g., medical devices and avionics, have critical
timing constraints.
In those settings, it can be catastrophic for the program to be forced to wait
for a significant period of time on the GC to finish collecting a region before
a new allocation can be made.

Until a memory limit is reached, state-of-the-art real-time garbage collection
algorithms can make the following guarantees~(see, e.g., Section~4.2~of~\citet{rtgc}):
\begin{enumerate}
    \item \textbf{Constant Pause Times:} The GC only slows each program
        instruction down by a constant amount.
    \item \textbf{Linear Collection Delay:} Suppose at some program point there
        are $n$ reachable memory regions, and then a program operation makes
        one of those regions unreachable. Then, the GC guarantees that region
        will be collected within $O(n)$ program operations.
\end{enumerate}
\textbf{The fundamental question of this paper is whether the $O(n)$ collection
delay can be reduced without increasing pause times.}
For na\"ive tracing GC algorithms, which are designed around a traversal of the
graph, the answer is intuitively no: no regions can be safely collected until
the entire traversal, which may have to visit $O(n)$ nodes, completes.
While the story is complicated by the use of incremental tracing, among GC
researchers, there is an informal understanding that the $O(n)$ collection
delay is unavoidable in some pathological scenarios, where the program may make
heap cycles and operates close to the memory limit.
However, until this paper, there was no formal proof ruling out the existence
of fundamentally more-clever algorithms yet to be discovered that improve
collection delay without worsening pause times.

\subsection{Problems with Delay}
Before explaining the theoretical results of our paper, it is helpful to
motivate our study by describing one application-level issue (GC thrashing) caused by the
existence of collection delay.
This issue is well-known among both researchers and
practitioners~\citep{yak,ejava,venners,jep}; we repeat it here merely to keep
the paper self-contained.
A more in-depth explanation is provided in~\pref{sec:Motivating}, and a
separate issue compounding on the already-problematic language feature of
finalization is described for interested readers in~\pref{app:MotivateFinalizers}.

\pref{sec:MotivateThrashing} describes a program operating very close to the memory
limit.
It first allocates a large amount of memory that stays reachable through the
entire program execution, pinning its logical memory usage near the limit.
It then repeatedly makes a single region unreachable before requesting a new
allocation in a loop.
Because of collection delay, modern GCs do not guarantee that the region just
made unreachable can be actually collected in time to be reused for the new
allocation.
Hence, to stay under the memory limit, the GC is forced to complete a full
collection cycle on each iteration, introducing linear-length pause times that
would be catastrophic for critical real-time applications.
In our example, the GC ends up thrashing and slows down end-to-end execution
time by $70\times$ compared to a manually managed version\footnote{In practice,
real-time systems based on collectors like that of~\citet{rtgc} work around
this issue by overprovisioning memory so the limit is never reached.
Overprovisioning incurs additional costs and is difficult to apply in settings
with dynamically changing memory constraints, e.g., with multiple interacting
processes.}.

\subsection{Impossibility Result}
It is tempting to hope that the issues described above and
in~\pref{sec:Motivating} could be fixed once-and-for-all with a more
complicated collector.
\textbf{The core result of this paper is an impossibility theorem implying it
is fundamentally impossible to avoid such issues~(\pref{cor:LB1}).}
For every garbage collector (defined formally in~\pref{sec:GCDS}) there is a
program where the GC must either (i)~introduce a superlogarithmic pause during
some program operation, or (ii)~delay collecting a region for a
superlogarithmic number of program operations.
If a long pause can be introduced, the collector is insufficient for use in
real-time settings (e.g., medical devices) where timing is critical.
On the other hand, if collection delay is introduced, the collector is
insufficient for applications that either (i) operate close to the memory
limit and so must be able to reuse memory regions quickly; or (ii) rely on
prompt finalization for actions like unlocking.

Notably, GC researchers have understood this fact for decades, but only at an
informal level.
To the best of our knowledge, ours is the first rigorous impossibility result
concerning garbage collection.
The result follows from a novel connection to the well-studied problem of
\emph{dynamic reachability} in the graph algorithms community.
Unfortunately, that connection goes both ways: although we expect the lower bound can
be improved (e.g., from superlogarithmic to linear), this is exactly as
difficult as solving a longstanding open problem in graph algorithms.

\subsection{Limitations of Results}
It is crucial to note a number of limitations to our results.
First, they only imply lower bounds for a GC algorithm insofar as the GC
algorithm can be used to implement our mutator-observer GCDS interface
in~\pref{sec:GCDS}.
Hence, they say nothing directly about, e.g., moving collectors.
However, it is sometimes possible to modify such GC algorithms to fit our GCDS
interface without significant effect on asymptotic time.
For example,~\citet{baker_treadmill} adapts the moving collector
in~\citet{copying_realtime} to a nonmoving version that can be used to implement
our GCDS interface.
In those cases, our bounds apply only to the modified algorithm, and additional
analysis must be done to understand to what extent (if at all) the results
imply anything about the unmodified algorithm.

Second, our results address only the problem of collecting regions no longer
reachable by pointers from a root set.
We say nothing about GC approaches using different approximations to liveness.

Finally, they imply merely that there exists \emph{at least one} pathological
sequence of pointer operations where the GC algorithm performs poorly, but they
say nothing about the performance of the algorithm in typical cases.
In particular, the pathological behavior we prove exists will involve operating
close to the memory limit and the existence of potential heap cycles.
Hence, our results are sidestepped when these pathological cases are ruled out
by restrictive type systems or resource overprovisioning, and they say nothing
about what can be achieved for the typical case by well-designed heuristics
like generational collection.

\subsection{Implications of Results}
\textbf{Our motivation is theoretical,} to understand with mathematical
certainty the limits of GC.
After nearly 60 years of research~\citep{ilp,gelernter_gc}, there still exist
pathological programs where even the best GCs significantly degrade
performance~(\pref{sec:Motivating}).
This paper provides a precise, formal explanation for the extent to which such
pathological cases are unavoidable, even by GC algorithms not yet discovered.

For GC researchers, our results provide precise and formal confirmation of the
informal understanding that searching for GC algorithms that avoid all such
pathological cases in the worst case is futile, hence motivating heuristics
like generational collection.
They also provide a helpful sanity check on claims; in~\pref{sec:Sanity} we
describe how one of our lower bounds would have helped discover a known error
in the cyclic reference counting literature.

\textbf{Our analysis has no direct implication for GC users,} beyond slightly
clarifying the worst-case tradeoffs that will come with the use of automated
GC.
In the very long term, we are optimistic that the algorithm described
in~\pref{sec:Algorithm} might spur research ideas that lead to better garbage
collectors, but our evaluation of the algorithm as presented in this paper
indicates that it is only beneficial in very extreme pathological scenarios,
hence not recommended for typical use.

\subsection{Contributions and Outline}
The major contributions of this paper are as follows:
\begin{enumerate}
    \item Definition of the \emph{Garbage Collection Data Structure} which
        formalizes the mutator--observer interface of most nonmoving
        garbage collection schemes~(\pref{sec:Background}).
    \item A novel lower bound showing fundamental tradeoffs between worst-case
        pause time and collection delay~(\pref{sec:LowerBounds1}).  It implies
        that every GC fitting the GCDS interface is susceptible to pathological
        cases similar to the ones demonstrated in~\pref{sec:MotivateThrashing}.
    \item A second lower bound showing that reference counting cannot be
        extended to handle cycles without either significantly increasing the
        worst-case pause times or introducing significant collection
        delay, even on programs making only acyclic heaps~(\pref{sec:LowerBounds2}).
    \item A GC guaranteeing $O(1)$ collection delay in all settings and
        logarithmic pause times while the heap is
        acyclic~(\pref{sec:Algorithm}). While it introduces too much overhead
        to suggest as a general-purpose GC, it addresses interesting and
        long-standing theoretical questions regarding the extent to which
        reference counting can be extended to handle cycles.
\end{enumerate}
\pref{sec:SummaryLB} ties the technical results back to the programming
languages context, \pref{sec:Future} discusses limitations and future work,
\pref{sec:Related} discusses related work, and \pref{sec:Conclusion} concludes
the paper.

\section{Motivating Example: What's Wrong with Delay?}
\label{sec:Motivating}
We motivate the question of whether garbage collectors with constant pause
times and collection delay exist using a well-known GC issue,
thrashing~\citep{yak}.
\pref{app:Motivating} has full code and explains how to trigger similar
behavior in GCs using generations, reference counting, etc.
In addition to the thrashing problem, collection delay is known to cause
deadlocks when combined with programs making overzealous use of finalizers.
However, the use of finalizers is generally known to be bad practice, so we
relegate an example of this to the appendix~(\pref{app:MotivateFinalizers}).

\begin{figure}[t]
\begin{minted}{Lua}
-- (excerpted implementations away of some functions, globals)
-- stage 1: fetch items into an array
-- each item has a large memory buffer associated with it
local items = {}
for i=1,N_ITEMS do items[i] = fetch_item() end
-- at this point, memory is near the limit
-- stage 2: compute a summary statistic
local summary = 0
-- process_item makes large, but temporary, allocations
-- frequent GC pauses needed to collect old temporary allocations
for i=1,N_ITEMS do summary = summary + process_item(items[i]) end
\end{minted}
    \caption{Excerpt from the example program showing long pause times in the
    memory-constrained setting.
    Both \texttt{fetch\_item} and \texttt{process\_item} allocate large memory
    blobs using a wrapper that enforces a strict limit on the total (logical)
    allocation size.
    The limit is almost reached by the end of stage 1, and only the blobs
    allocated in stage 2 are temporary, so nearly every iteration of the loop
    in stage 2 needs to make a full GC pass to collect old temporary blobs
    before allocating a new one for that iteration.
    This program is sufficient to trigger GC thrashing in Lua.
    Variants of this program trigger similar linear-length pause times in
    collectors that use generational collection and reference counting.
    }
    \label{fig:MotivatingMemoryListing}
\end{figure}

\subsection{GC Thrashing and Linear Pause Times in the Memory-Constrained Setting}
\label{sec:MotivateThrashing}
Real-time applications critically require constant-length worst-case pause
times.
Existing garbage collection schemes can guarantee constant-length pause times.
But because they do not guarantee immediate collection, when a program operates
close to the memory limit---common in, e.g., embedded settings where real-time
guarantees are important---existing GCs can be forced to choose between
breaking this guarantee and waiting significantly longer to complete a
collection cycle before enough free memory can be found to continue execution,
or refusing an allocation even though unreachable memory does exist.

Demonstrating this, consider the Lua program excerpted
in~\pref{fig:MotivatingMemoryListing}.
The calls to \texttt{fetch\_item} and \texttt{process\_item} both allocate
large memory buffers.
The program attempts to limit overall memory usage via a small wrapper (not
shown in the excerpt) that records the total number of uncollected allocations
and refuses to allocate more if the limit is reached.

The application's first stage fills an array of objects.
Each object contains a reference to a large memory blob, e.g., the contents of
a file.
These blobs will stay permanently reachable throughout the program execution.
After the first stage we have nearly reached our memory limit.

The second stage computes a summary, e.g., the most frequent token in the file,
from each item in the array.
Computing the summary requires allocating another large, but temporary, memory
buffer for each item.
Hence, the memory limit gets reached quickly after the first few iterations of
the second stage.
Seeing this, the allocator must wait for the GC to finish collecting old
temporary buffers from previous iterations before it can allocate space for the
latest iteration of the second loop.
This causes repeated waiting on the GC, often referred to as \emph{GC
thrashing}.

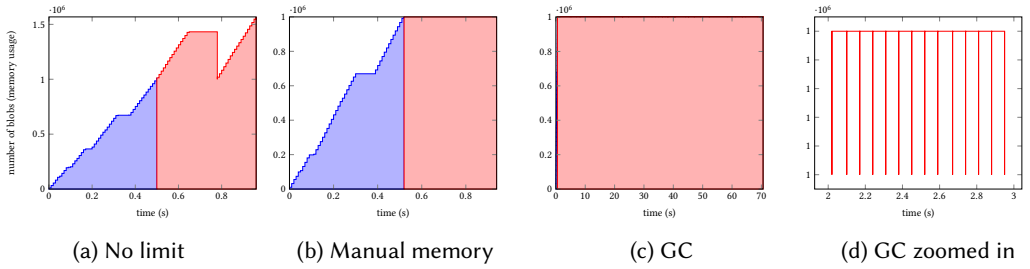
\begin{figure}[t]
    \centering
    \begin{subfigure}{0.24\textwidth}
        \begin{tikzpicture}[scale=0.4]
            \begin{axis}[xlabel=time (s),ylabel=number of blobs (memory usage),area style,enlargelimits=false]
                \addplot+[no markers] table[x index=0,y index=1,col sep=comma] {data/pressure.nolimit.stage_0.csv} \closedcycle;
                \addplot+[no markers] table[x index=0,y index=1,col sep=comma] {data/pressure.nolimit.stage_1.csv} \closedcycle;
            \end{axis}
        \end{tikzpicture}
        \caption{No limit}
        \label{fig:MotivatingPressure1}
    \end{subfigure}
    \hfill
    \begin{subfigure}{0.24\textwidth}
        \begin{tikzpicture}[scale=0.4]
            \begin{axis}[xlabel=time (s),area style,enlargelimits=false]
                \addplot+[no markers] table[x index=0,y index=1,col sep=comma] {data/pressure.manual.stage_0.csv} \closedcycle;
                \addplot+[no markers] table[x index=0,y index=1,col sep=comma] {data/pressure.manual.stage_1.csv} \closedcycle;
            \end{axis}
        \end{tikzpicture}
        \caption{Manual memory}
        \label{fig:MotivatingPressure2}
    \end{subfigure}
    \hfill
    \begin{subfigure}{0.24\textwidth}
        \begin{tikzpicture}[scale=0.4]
            \begin{axis}[xlabel=time (s),area style,enlargelimits=false]
                \addplot+[no markers] table[x index=0,y index=1,col sep=comma] {data/pressure.limit.stage_0.csv} \closedcycle;
                \addplot+[no markers] table[x index=0,y index=1,col sep=comma] {data/pressure.limit.stage_1.csv} \closedcycle;
            \end{axis}
        \end{tikzpicture}
        \caption{GC}
        \label{fig:MotivatingPressure3}
    \end{subfigure}
    \hfill
    \begin{subfigure}{0.24\textwidth}
        \begin{tikzpicture}[scale=0.4]
            \begin{axis}[xlabel=time (s),enlargelimits=true]
                \addplot+[red,no markers,x filter/.expression={(and(x >= 2, x <= 3) ? \pgfmathresult : NaN)}] table[x index=0,y index=1,col sep=comma] {data/pressure.limit.stage_1.csv};
            \end{axis}
        \end{tikzpicture}
        \caption{GC zoomed in}
        \label{fig:MotivatingPressure4}
    \end{subfigure}
    \caption{With and without memory pressure. The first stage is shown in
    blue, the second in red.}
    \label{fig:MotivatingPressure}
\end{figure}

Logical memory usage is graphed against time in~\pref{fig:MotivatingPressure}.
Without a memory limit~(\pref{fig:MotivatingPressure1}) the GC is never forced
to run, so the application finishes quickly with only a small number of
relatively short pauses.
When a limit is enforced but the programmer manually marks unreachable memory
for collection~(\pref{fig:MotivatingPressure2}), the limit is never reached so
performance is similarly good.

However, when the memory limit is enforced and the GC is relied on to
automatically reclaim unreachable regions~(\pref{fig:MotivatingPressure3}), we
see a $70\times$ slowdown.
Nearly all of this time is spent in the second stage thrashing within the GC.
\pref{fig:MotivatingPressure4} shows a zoomed-in view of one second's worth of
the second stage execution.
The majority of time is spent with memory usage pinned at the maximum while the
GC is running to free up new space for a new allocation.
The actual application work only happens within the near instantaneous dips
below that limit.

Note that all of the memory allocated during the first stage is still
reachable, so the GC cannot collect more than a small amount at a time.
Specifically, it collects temporary regions from calls to
\texttt{process\_item} since the last time it was run.
Also note that the time taken by the GC is not time needed to actually free the
memory, as the manual memory management performance is significantly better.
Instead, the time is spent \emph{locating what can be freed}: ensuring even a
single region can be freed requires searching the \emph{entire} heap to check
that nothing else points to it.

This pathological behavior shows one downside of collection delay: when more
memory is needed, there is no guarantee that the GC will be able to find it
quickly even if it exists.
Our impossibility result proves that this sort of scenario is impossible to
avoid.

\section{Preliminaries and the Garbage Collection Data Structure}
\label{sec:Background}
To rigorously investigate claims about what is or is not possible for future
garbage collectors, we must formalize what we mean by a garbage collector.
This section formalizes the garbage collection problem as the garbage
collection data structure (GCDS) and discusses nuances in defining the
asymptotic running time of GCDS operations.

\subsection{Garbage Collection Data Structure (GCDS)}
\label{sec:GCDS}
We will model runtime pointer information as a directed multigraph, i.e., a set
$V$ of vertices (nodes) and a map $E : V \times V \to \mathbb{N}_{\geq 0}$
indicating the number of edges from one node to another.
Vertices represent memory regions and the edges represent pointers from one
region to another (we will use ``regions'' and ``nodes'' interchangeably
throughout).
Because we are proving \emph{lower} bounds, we are justified in ignoring the
size of memory regions; the worst-case sequence of operations we will prove
exists can be thought of as making equal-sized allocations.

A \emph{GC heap} is a directed multigraph along with a distinguished vertex
$\GCRoot \in V$ with no incoming edges. The $\GCRoot$ vertex is meant to
represent the local and global variables, i.e., it has outgoing edges to
any regions that local and global variables have direct pointers to.

With this in mind, we can define a set of procedures that describe the desired
interface between the programming language and the garbage collector.
The interface was designed to capture the standard interface of nonmoving
GCs for imperative code, i.e., where the GC sees the program as a mutator
modifying pointers in a heap.
\begin{definition}
    \label{def:GCDS}
    A \emph{garbage collection data structure} (GCDS) is a data structure that
    (i) represents a GC heap, i.e., a directed multigraph with distinguished
    vertex $\GCRoot$; (ii) stores a list $\GCFreeList$ of vertices not
    reachable from $\GCRoot$; and (iii) supports the following procedures:
    \begin{itemize}
        \item $\GCAllocate()$ allocates a new vertex $a$ and adds edge $\GCRoot
            \to a$.
        \item $\GCInsert(a \to b)$ adds edge $a \to b$.
        \item $\GCDelete(a \to b)$ removes edge $a \to b$.
        \item $\GCStep()$ requests the GCDS perform additional collection work
            (analogous to, e.g., \\ \texttt{collectgarbage("step")} in Lua).
    \end{itemize}
    The GCDS may only add nodes that are unreachable from $\GCRoot$ to
    $\GCFreeList$.
    While the user of the data structure may remove nodes from $\GCFreeList$,
    the GCDS itself must never remove a node from $\GCFreeList$ (only add).
    The user of the data structure promises no edge ending in $\GCRoot$ is
    inserted and no query involves a node no longer reachable from $\GCRoot$.
    We say each call made to one of these GCDS procedures is a \emph{GCDS
    operation}.
\end{definition}

A GCDS can be used to implement GC for an imperative language.
When the program overwrites a pointer in region~$s$ that was pointing to
region~$d_1$ to now point to region~$d_2$, the GCDS operations~$\GCInsert(s \to
d_2)$ and~$\GCDelete(s \to d_1)$ are performed.
$\GCStep()$ can be called after every program operation to perform some
incremental GC work.
When memory pressure is encountered, $\GCFreeList$ can be checked for
allocation regions that can be reused by the application.
If $\GCFreeList$ is empty and the application requires more memory, $\GCStep()$
can be called to request the collector spend more time searching for regions
that may be safely freed.
The extent to which an efficient GC implies the existence of an efficient GCDS
is discussed in~\pref{app:GC2GCDS}.

Note that the collector does \emph{not} need to ensure that all unreachable
nodes are placed on $\GCFreeList$ immediately upon becoming unreachable, or
even after a call to $\GCStep()$.
This allows for modelling concurrent tracing collectors that perform only a
small portion of a sweep after each program operation.
The frequency at which $\GCFreeList$ is updated to reflect newly unreachable
nodes is captured by the \emph{delay} metric below~(\pref{def:Delay}).

\subsection{Defining Asymptotic Complexity of Garbage Collection}
\label{sec:DefiningTime}
We consider two primary measures of GCDS efficiency.
These definitions are somewhat informal for space reasons; more formal
definitions and a discussion of alternative definitions that also work is given
in~\pref{app:FormalDefinitions}.
We will assume here the GCDS is deterministic.

\subsubsection{Worst-Case Delay}
The focus of this paper is on \emph{worst-case delay}, i.e., the number of GC
operations that may need to be executed before an unreachable node is added to
$\GCFreeList$.
\begin{definition}
    \label{def:Delay}
    The \emph{worst-case delay} $d(n)$ of a GCDS is the pointwise minimal
    function such that, in any sequence of GCDS operations involving at most
    $n$ nodes (i.e., making at most $n$ calls to $\GCAllocate$), if some
    operation makes a node $u$ unreachable, $u$ is added to $\GCFreeList$ after
    at most $d(n)$ more GCDS operations (including the operation that makes it
    unreachable).
\end{definition}

\subsubsection{Worst-Case Pause Times}
The other quantity we are interested in is the \emph{worst-case pause time},
i.e., the longest amount of time that any single GCDS operation might take.
This corresponds directly to the extra pause time between program operations
introduced by the GC.
In a real-time application, guaranteeing $O(1)$ pause times can be critical.

\begin{definition}
    \label{def:FullTime}
    The \emph{worst-case pause time} $t(n)$ of a GCDS is the pointwise
    minimal function such that, in any sequence of GCDS operations involving at
    most $n$ nodes, each GCDS operation in the sequence takes time at most
    $t(n)$.
\end{definition}

\subsubsection{Fine-Grained Measures: Noncollecting and Acyclic Pause Times}
We will eventually prove that, for every GCDS, there is a sequence of
operations that forces it to introduce either a long pause time or a long
collection delay.
The reader may be concerned that the sequence of operations is somehow
`unfair,' e.g., that long pause times are only encountered because a large
fraction of the heap needs to be collected quickly, or that densely connected
structures are used, or that cyclic structures are used.
To address these concerns we will need to introduce the \emph{all-reachable},
\emph{acyclic all-reachable}, and \emph{sparse acyclic all-reachable} variants
of the worst-case pause time definition from earlier.
Each one captures the worst-case pause times introduced for a subset of
possible program executions.
Hence, lower bounds on these are stronger than lower bounds on $t(n)$, as they
guarantee the existence of a hard sequence of operations involving
progressively simpler classes of heaps.
\begin{definition}
    \label{def:ARTime}
    The \emph{all-reachable} worst-case pause time $\tAR(n)$ of a GCDS is the
    pointwise minimal function such that, in any sequence of GCDS
    operations involving at most $n$ nodes \textbf{and never making any node
    unreachable}, each operation in the sequence takes time at most $\tAR(n)$.
\end{definition}
\begin{definition}
    \label{def:AARTime}
    The \emph{acyclic all-reachable} worst-case pause time $\tAAR(n)$ of a
    GCDS is the pointwise minimal function such that, in any sequence of
    GCDS operations involving at most $n$ nodes, \textbf{never making any node
    unreachable, and never forming a cycle in the heap}, each operation in
    the sequence takes time at most $\tAAR(n)$.
\end{definition}
\begin{definition}
    \label{def:SAARTime}
    The \emph{sparse acyclic all-reachable} worst-case pause time $\tSAAR(n)$
    of a GCDS is the pointwise minimal function such that, in any
    sequence of GCDS operations involving at most $n$ nodes, \textbf{never
    making any node unreachable, never forming a cycle in the heap, and never
    having more than $O(1)$ edges leaving any node}, each operation in the
    sequence takes time at most $\tSAAR(n)$.
\end{definition}

It is important to clearly note here that lower bounds on these restricted
versions of $t(n)$ are \emph{stronger} than lower bounding $t(n)$ itself.
\begin{lemma}
    \label{lem:Ts}
    Every GCDS has $\tSAAR(n) \leq \tAAR(n) \leq \tAR(n) \leq t(n)$.
\end{lemma}
\begin{proof}
    Every function in the sequence is defined as the max time over
    progressively larger subsets of possible executions.
    The claim follows because the max over a subset is smaller than that over
    the full set, i.e., when $S \subseteq T$ we know $\mathrm{max}(S) \leq
    \mathrm{max}(T)$.
\end{proof}

\subsection{Examples of GCDS Implementations}
\label{sec:GCDSExamples}
Our theoretical results~(\pref{sec:LowerBounds1} and~\pref{sec:LowerBounds2})
apply to any GC approach that can be adapted to implement the GCDS
interface~(\pref{def:GCDS}).
It is therefore important to get a sense for how different GC approaches can
be adapted to implement the GCDS interface.
This section describes how to use both reference counting and mark-and-sweep to
implement the GCDS interface.

In the below pseudocode, we make use of multisets to store the edges in the
heap graph.
We assume multisets are implemented such that the following can be done in
$O(1)$ time:
\begin{enumerate}
    \item They can be converted into an iterator that only visits each element
        once; we call this operation $s.\mathrm{no\_multi}()$.
    \item The number of copies of an object in the multiset can be computed
        using $s.\mathrm{count}(o)$.
\end{enumerate}

\subsubsection{Eager Reference Counting as a GCDS}
\begin{figure}
    \begin{minipage}[t]{0.48\textwidth}
        \begin{algorithm}[H]
            \caption{$\GCInit()$} \label{alg:rc_gcinit}
            $\mathrm{ref\_counts} \gets \emptyset$\;
            $\mathrm{outgoing} \gets \emptyset$\;
            $\GCRoot \gets \mathrm{id} \gets 0$\;
        \end{algorithm}
        \begin{algorithm}[H]
            \caption{$\GCStep()$} \label{alg:rc_gcstep}
            \textbf{no-op}\;
        \end{algorithm}
    \end{minipage}
    \hfill
    \begin{minipage}[t]{0.48\textwidth}
        \begin{algorithm}[H]
            \caption{$\GCAllocate()$} \label{alg:rc_gcallocate}
            $\mathrm{id} \gets \mathrm{id} + 1$\;
            \Return{$\mathrm{id}$}\;
        \end{algorithm}
        \begin{algorithm}[H]
            \caption{$\GCInsert(a \to b)$} \label{alg:rc_gcinsert}
            $\mathrm{ref\_counts}[b] \gets \mathrm{ref\_counts}[b] + 1$\;
            $\mathrm{outgoing}[a] \gets \mathrm{outgoing}[a] \cup \{ b \}$\;
        \end{algorithm}
    \end{minipage}
\begin{algorithm}[H]
    \caption{$\GCDelete(a \to b)$} \label{alg:rc_gcdelete}
    $\mathrm{outgoing}[a] \gets \mathrm{outgoing}[a] - \{ b \}$\;
    $\mathrm{ref\_counts}[b] \gets \mathrm{ref\_counts}[b] - 1$\;
    \If{$\mathrm{ref\_counts}[b] = 0$}{
        $\mathrm{worklist} \gets \{ b \}$\;
        \While{$\mathrm{worklist} \neq \emptyset$}{
            $\mathrm{node} \gets \mathrm{worklist}.\mathrm{pop}()$\;
            $\GCFreeList{} \gets \GCFreeList{} \cup \{ \mathrm{node} \}$\;
            \For{$c \in \mathrm{outgoing}[\mathrm{node}].\mathrm{no\_multi}()$}{
                $\mathrm{ref\_counts}[c] \gets \mathrm{ref\_counts}[c] - \mathrm{outgoing}[\mathrm{node}].\mathrm{count}(c)$\;
                \lIf{$\mathrm{ref\_counts}[c] = 0$}{$\mathrm{worklist} \gets \mathrm{worklist} \cup \{ c \}$}
            }
        }
    }
\end{algorithm}
    \caption{Eager reference counting as an implementation of the GCDS interface. ID
    0 represents the root node (stack/local/global variables) while all other
    memory regions get a unique ID upon allocation. \texttt{ref\_counts} is a
    map from IDs to counts that could be implemented using a hashmap.}
    \label{fig:RCGCDS}
\end{figure}

\pref{fig:RCGCDS} shows an implementation of the GCDS interface using the
eager reference counting technique.
The GCDS's internal data includes a dictionary mapping each node ID to a
reference count, a multiset of outgoing edges for each node ID, and a running
ID used to allocate new node IDs.

The $\GCRoot$ node is associated with ID zero.
Allocation involves assigning a new ID and returning it for use to the user.
Inserting an edge $a \to b$ updates the reference count of $b$ and outgoing
edge set of $a$.
Deleting an edge $a \to b$ reduces the reference count of $b$; if it drops to
zero, $b$ is placed on the free list and all of its outgoing edges are
similarly deleted in a recursive manner.

What is important in this context is not that the GCDS perfectly match the
implementation of a real reference counting system (indeed, this would require
specializing to a specific language implementation and runtime, which we are
trying to abstract away from), but rather ensuring that the performance
characteristics we defined earlier (delay and pause time) are not harmed in the
translation from real GC approach to GCDS implementation.

Notably, this GCDS does not attempt to store reference counts within the
allocated region itself; in fact, it has no notion of "allocated memory" at all
because it operates on a graph abstraction of the heap.
Intuitively, one can think of the GCDS as operating in a separate address space
from the program, seeing only pointer operations as sequences of GCDS
operations on handles that are returned by $\GCAllocate$.
Hence, it must store its own metadata and shadow copy of the graph, e.g., using
the `$\mathrm{ref\_counts}$' and `$\mathrm{outgoing}$' variables
in~\pref{fig:RCGCDS}.
However, it is important to note that searching in these variables can be done
in $O(1)$ expected time using a hashmap to store the refcounts and outgoing
edges, or in $O(1)$ worst-case time using a direct addressing table at the
expense of added memory overhead.
In general, our impossibility results in this paper apply \emph{regardless of
the space usage of the GCDS itself}, hence it is okay if when translating a GC
algorithm to implement the GCDS interface more space (even asymptotically more
space) is used, as long as the delay and pause times are not affected;
all of our results will still apply.

\textbf{Worst-Case Delay:} This reference counting GCDS has no well-defined
worst-case delay, because if a sequence of GCDS operations forms a cycle, e.g.,
$\GCRoot \to a \to b \to c \to a$, and then deletes $\GCRoot \to a$, the node
$a$ will \emph{never} be added to the free list, no matter how many additional
GC operations are performed. We informally write $d(n) = \infty$ for this
situation.

\textbf{Worst-Case Pause Times:}
In some cases, a single edge removal could make every node in the heap
unreachable at once resulting in $\GCDelete$ having to iterate over every edge
in the heap, making $t(n) = O(n^2)$.
However, the worst-case \emph{all-reachable} pause time (\pref{def:ARTime}) is
\emph{significantly} better: if all nodes are still reachable from $\GCRoot$,
then the reference count could not have dropped to zero, and hence the if
condition in $\GCDelete$ is never taken.
All the remaining operations take $O(1)$ time, hence $\tAR(n) = O(1)$.
By~\pref{lem:Ts} this also implies $\tAAR(n) = O(1)$ and $\tSAAR(n) = O(1)$.

\subsubsection{Mark-and-Sweep as a GCDS}
\begin{figure}
    \begin{minipage}[t]{0.48\textwidth}
        \begin{algorithm}[H]
            \caption{$\GCInit()$} \label{alg:mas_gcinit}
            $\mathrm{outgoing} \gets \emptyset$\;
            $\GCRoot \gets \mathrm{id} \gets 0$\;
            $\mathrm{all} \gets \emptyset$
        \end{algorithm}
        \begin{algorithm}[H]
            \caption{$\GCInsert(a \to b)$} \label{alg:mas_gcinsert}
            $\mathrm{outgoing}[a] \gets \mathrm{outgoing}[a] \cup \{ b \}$\;
        \end{algorithm}
    \end{minipage}
    \hfill
    \begin{minipage}[t]{0.48\textwidth}
        \begin{algorithm}[H]
            \caption{$\GCAllocate()$} \label{alg:mas_gcallocate}
            $\mathrm{id} \gets \mathrm{id} + 1$\;
            $\mathrm{all} \gets \mathrm{all} \cup \{\mathrm{id}\}$\;
            \Return{$id$}\;
        \end{algorithm}
        \begin{algorithm}[H]
            \caption{$\GCDelete(a \to b)$} \label{alg:mas_gcdelete}
            $\mathrm{outgoing}[a] \gets \mathrm{outgoing}[a] - \{ b \}$\;
            $\GCStep()$\;
        \end{algorithm}
    \end{minipage}
    \begin{algorithm}[H]
        \caption{$\GCStep()$} \label{alg:mas_gcstep}
        $\mathrm{worklist} \gets \{ \GCRoot \}$\;
        $\mathrm{marked} \gets \emptyset$\;
        \While{$\mathrm{worklist} \neq \emptyset$}{
            $\mathrm{node} \gets \mathrm{worklist}.\mathrm{pop}()$\;
            \lIf{$\mathrm{node} \in \mathrm{marked}$}{\textbf{Continue}}
            $\mathrm{marked} \gets \mathrm{marked} \cup \{ \mathrm{node} \}$\;
            $\mathrm{worklist} \gets \mathrm{worklist} \cup \mathrm{outgoing}[\mathrm{node}].\mathrm{no\_multi}()$\;
        }
        $\GCFreeList \gets \GCFreeList \cup (\mathrm{all} - \mathrm{marked})$\;
        $\mathrm{all} \gets \mathrm{all} - (\mathrm{all} - \mathrm{marked})$\;
    \end{algorithm}
    \caption{Nonincremental mark-and-sweep as an implementation of the GCDS
    interface. ID 0 represents the root node (stack/local/global variables)
    while all other memory regions get a unique ID upon allocation.
    \texttt{outgoing} maps each IDs to a multiset of IDs, while
    \texttt{worklist} is a non-multi set.}
    \label{fig:MASGCDS}
\end{figure}

\pref{fig:MASGCDS} shows how the mark-and-sweep technique can be used to
implement the GCDS interface.
Like the reference counting GCDS, we keep track of a shadow copy of the heap
graph.
But instead of checking reference counts to determine when something becomes
unreachable, we perform a search through the graph starting at $\GCRoot$ to
determine any newly unreachable nodes.

\textbf{Worst-Case Delay:}
This GCDS has worst-case delay $d(n) = 1$, because it guarantees that all
unreachable regions are added to the free list \emph{immediately} once they
become unreachable.

\textbf{Worst-Case Pause Times:}
This GCDS has worst-case pause time $t(n) = O(n^2)$ because the marking phase
in $\GCDelete$ sometimes has to visit every node in the graph, and for each of
those nodes must add all of its outgoing nodes to the worklist.
This analysis is unaffected by the connectedness or cyclicity of the heap
graph, hence we have $\tAR(n) = O(n^2)$ and $\tAAR(n) = O(n^2)$.
But if the heap graph is sparse, then the number of edges is limited and we get
$\tSAAR(n) = O(n)$.

\subsection{Cell Probe Model, Random Access Machines, and Reductions}
\label{sec:CPM}
The existing lower bounds we use~\citep{lbdc,bestlbdc} are in the cell probe model
of~\citet{cpm}. Data structures in the cell probe model are split into a single
\emph{persistent store} and a set of \emph{procedures}. The persistent store is
a large table of binary words, analogous to the random access memories used in
modern computers.  The procedures are programs that can read and write words from
the persistent store.

Lower bounds in the cell probe model bound only the
number of reads from and writes to the persistent store.
No assumption at all is made about the machine model that the procedures run
on, except that accessing a memory cell takes $\Omega(1)$ time and that
writes made to the store are a deterministic function of reads from the store.
In fact, lower bounds proved in this way apply even to unrealistic machine
models, e.g., Turing machines with an oracle for the halting problem.

One nuance in the cell probe model is the need for a word size for the table.
Usually, the data structure is given an upper bound $\NMax$ on the number of
objects the data structure might be asked to represent, e.g., nodes and edges
in a graph or items in a set. On a realistic modern machine,~$\NMax$~would be
approximately $2^{64}$.  The persistent store is allowed to have a word size
logarithmic in this upper bound, i.e.,~$w = O(\log{\NMax})$.

\section{Main Lower Bound}
\label{sec:LowerBounds1}
Our goal in this section is to prove that every GCDS has $t(n)d(n) =
\tilde{\Omega}(\log^{3/2}{n})$.
In particular, this means any GCDS guaranteeing $O(1)$ delay must have
superlogarithmic pause times.
In fact, we will prove the stronger result that $\tAR(n)d(n) =
\tilde{\Omega}(\log^{3/2}{n})$, i.e., the superlogarithmic pause can be
triggered by a sequence of GCDS operations where nothing ever becomes
unreachable.
This is, at first glance, counterintuitive because if it knew ahead of time
that nothing becomes unreachable, the GCDS would not need to do any work.
However, the GCDS is not told this ahead of time, and we still require that it
guarantee delay $d(n)$ if something \emph{were} to become unreachable.
Essentially, the GCDS must still prove to itself that nothing has become
unreachable.

\subsection{Dynamic Graph Reachability}
Our main lower bound follows via reduction from dynamic graph reachability,
defined below.
\begin{definition}
    \label{def:DRDS}
    A \emph{dynamic reachability data structure} (DRDS) stores a graph $(V,
    E)$ on a fixed number $\NMax = n$ of vertices. It supports the following
    operations:
    \begin{enumerate}
        \item $\DCInsert(a \to b)$ adds edge $a \to b$.
        \item $\DCDelete(a \to b)$ removes edge $a \to b$.
        \item $\DCConnected(a \to^+ b)$ returns \emph{reaches} if and only if
            there is a path from $a$ to $b$.
    \end{enumerate}
\end{definition}

\citet{bestlbdc} prove that this general form of dynamic reachability
requires~$\tilde{\Omega}(\log^{3/2}{n})$ time in the cell probe
model~(\pref{sec:CPM})~\citep{cpm}.
\begin{theorem} (\citet{bestlbdc})
    Every DRDS has worst-case~$\tilde{\Omega}(\log^{3/2}{n})$ per-operation
    time.
\end{theorem}

\subsection{Clocked Machines and Checkpoint-Restore}
\label{sec:FancyMachines}
Our reduction involves two special operations: \emph{checkpoint-restore} and
\emph{timeouts}.

A `restore' operation causes the state of the data structure's persistent store
to be restored to that of the last `checkpoint' operation.
Finitely many checkpoint-restore operations can be implemented with constant
overhead on realistic machines by logging memory writes after each checkpoint
and undoing them when restoring.

Timeouts stop the operation of a subroutine after a fixed number of
instructions are executed.
They can be implemented with $O(1)$ overhead by instrumenting the subroutine
instructions to continuously increment a clock counter and exit the subroutine
if the counter passes the timeout.

\subsection{Reduction from~\citet{bestlbdc}, DRDS}
\label{sec:LB1}
We now show that an efficient GCDS could be used to construct an efficient
DRDS.
The existence of such a reduction means that the known lower bounds on the DRDS
problem apply to GCDS as well.
The reduction is described by~Algorithms~\ref{alg:lbinsert}--\ref{alg:lbquery}.

\begin{figure}[t]
    \begin{minipage}{0.52\textwidth}
        \begin{algorithm}[H]
            \caption{$\DCInsert(v_i \to v_j)$} \label{alg:lbinsert}
            $v_i, v_j \gets \DCCreate(i, j)$\;
            $\GCInsert(v_i \to v_j)$\;
        \end{algorithm}
        \vspace{4mm}
        \begin{algorithm}[H]
            \caption{$\DCDelete(v_i \to v_j)$} \label{alg:lbdelete}
            $v_i, v_j \gets \DCCreate(i, j)$\;
            $\GCDelete(v_i \to v_j)$\;
        \end{algorithm}
    \end{minipage}
    \hfill
    \begin{minipage}{0.45\textwidth}
        \begin{algorithm}[H]
            \caption{$\DCCreate(i_1, i_2, \ldots)$} \label{alg:lbinit}
            \lIf{$X = \bot$}{
                $X \gets \GCAllocate()$
            }
            \For{$i_k \in i_1, i_2, \ldots$}{
                \If{$i_k \not\in N$}{
                    $N(i_k) \gets \GCAllocate()$\;
                    $\GCInsert(X \to N(i_k))$\;
                    $\GCDelete(\GCRoot \to N(i_k))$\;
                }
            }
            \Return{$N(i_1), N(i_2), \ldots$}\;
        \end{algorithm}
    \end{minipage}
    \begin{algorithm}[H]
        \caption{$\DCConnected(v_i \to^+ v_j)$} \label{alg:lbquery}
        \textbf{\textsc{Checkpoint}}\;
        $v_i, v_j \gets \DCCreate(i, j)$\;
        $\GCInsert(\GCRoot \to v_i, v_j \to X)$\;
        \tcc{The GCDS guarantees that operations take $\leq \tAR(n)$ time when
        nothing new becomes unreachable, so if it runs longer than that we know
        something must have become unreachable.}
        \If{$\GCDelete(\GCRoot \to X)$ runs for more than $\tAR(n)$ instructions or $\GCFreeList$ is not empty}{
            \textbf{\textsc{Restore}}\;
            \Return{\textsc{NotReaches}}\;
        }
        \tcc{The GCDS guarantees that any unreachable region is collected
        within $d(n)$ operations, so this loop simulates $d(n)$ program
        operations to guarantee that unreachable regions will be detected if
        any exist.}
        \ForEach{$i \in 1, 2, \ldots, d(n)$}{
            \If{$\GCStep()$ runs for more than $\tAR(n)$ instructions or $\GCFreeList$ is not empty}{
                \textbf{\textsc{Restore}}\;
                \Return{\textsc{NotReaches}}\;
            }
        }
        \textbf{\textsc{Restore}}\;
        \Return{\textsc{Reaches}}\;
    \end{algorithm}
    \caption{Reduction from DRDS to GCDS. The $\DCCreate$ helper method is
    needed to lazily create nodes in the GCDS because the DRDS comes
    preallocated with $n$ nodes while the GCDS begins with none.}
    \label{fig:LBAlgs}
\end{figure}

\paragraph{High-Level Operation of the Reduction.}
At a high level, we implement the DRDS operations $\DCInsert$ and $\DCDelete$
by mirroring their edge manipulations in the GCDS.
The GCDS also sees an additional node, $X$, that has an incoming edge from
$\GCRoot$ and an outgoing edge to every other node in the graph.
To check for the existence of a path between two nodes, we connect $\GCRoot$ to
the source, connect the destination to~$X$, and then disconnect $X$ from the
root.
There was a path from source to destination if and only if all nodes remain
reachable.

\paragraph{All-Reachable Time Assumptions.}
Recall our goal is to show the stronger fact that $\tAR(n)d(n) =
\tilde{\Omega}(\log^{3/2}{n})$, so our reduction we may assume \emph{only} that
the all-reachable worst-case pause times are guaranteed to be low.
The GCDS is allowed to take an arbitrarily long time---or even refuse to
terminate---if there are unreachable nodes.
How should we use the GCDS operations, then, if they might cause things to
become unreachable?
An elegant solution is to timeout the execution of the
GCDS~(\pref{sec:FancyMachines}): if it takes more than $\tAR(n)$ steps, it
\emph{must} indicate there are unreachable regions.
Otherwise, we use its output to determine whether there are any unreachable
nodes.

\paragraph{Destructive Operations.}
The $\DCConnected$ reduction requires calling $\GCInsert$ and $\GCDelete$,
which modify the heap, even though $\DCConnected$ is a read-only operation. To
address this, we checkpoint and restore the data
structure~(\pref{sec:FancyMachines}) so the persistent store does not see any
writes performed during the call to $\DCConnected$.

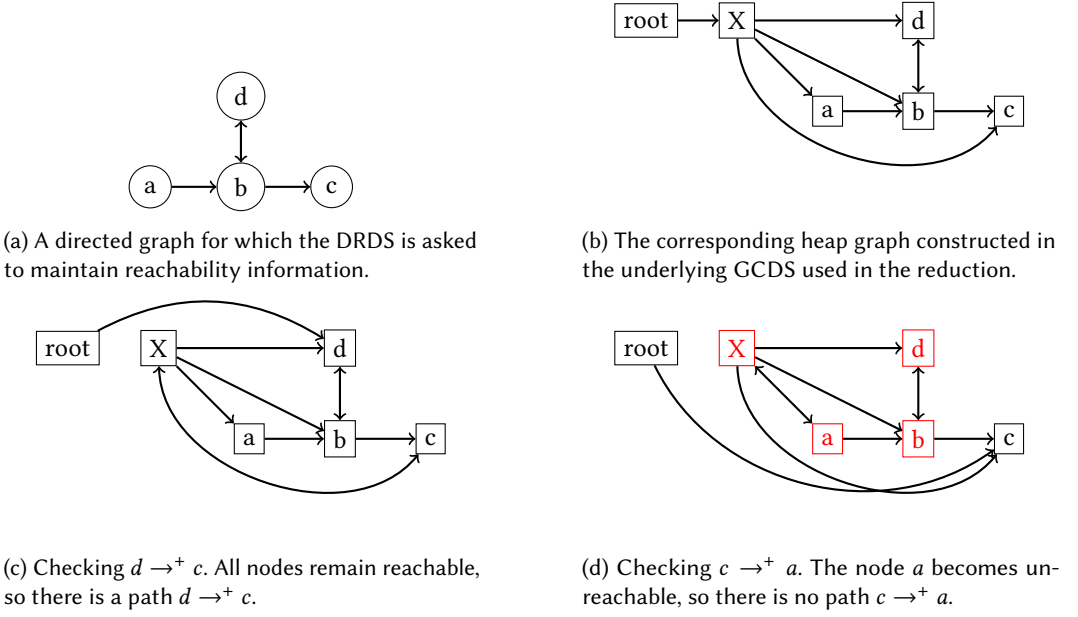
\begin{figure}[t]
    \centering
    \begin{subfigure}[t]{0.45\textwidth}
        \center
        \begin{tikzpicture}[scale=0.8]
            \node[gnode] (a) at (0, 0) {a};
            \node[gnode] (b) at (1.5, 0) {b};
            \node[gnode] (c) at (3, 0) {c};
            \node[gnode] (d) at (1.5, 1.5) {d};
            \draw[thick] (a) edge[->] (b) (b) edge[->] (c);
            \draw[thick] (b) edge[<->] (d);
        \end{tikzpicture}
        \caption{A directed graph for which the DRDS is asked to maintain
        reachability information.}
        \label{fig:lb1A}
    \end{subfigure}
    \hfill
    \begin{subfigure}[t]{0.45\textwidth}
        \center
        \begin{tikzpicture}[scale=0.8]
            \node[gcnode] (root) at (-3, 1.5) {root};
            \node[gcnode] (X) at (-1.5, 1.5) {X};

            \node[gcnode] (a) at (0, 0) {a};
            \node[gcnode] (b) at (1.5, 0) {b};
            \node[gcnode] (c) at (3, 0) {c};
            \node[gcnode] (d) at (1.5, 1.5) {d};

            \draw[thick] (a) edge[->] (b) (b) edge[->] (c);
            \draw[thick] (b) edge[<->] (d);

            \draw [->,thick] (root) -- (X);
            \draw [->,thick] (X) -- (a);
            \draw [->,thick] (X) -- (b);
            \draw [->,thick] (X) -- (d);
            \draw [->,thick] (X) to[bend right=70] (c);
        \end{tikzpicture}
        \caption{The corresponding heap graph constructed in the underlying GCDS
        used in the reduction.}
        \label{fig:lb1B}
    \end{subfigure}
    \begin{subfigure}{0.45\textwidth}
        \center
        \begin{tikzpicture}[scale=0.8]
            \node[gcnode] (root) at (-3, 1.5) {root};
            \node[gcnode] (X) at (-1.5, 1.5) {X};

            \node[gcnode] (a) at (0, 0) {a};
            \node[gcnode] (b) at (1.5, 0) {b};
            \node[gcnode] (c) at (3, 0) {c};
            \node[gcnode] (d) at (1.5, 1.5) {d};

            \draw[thick] (a) edge[->] (b) (b) edge[->] (c);
            \draw[thick] (b) edge[<->] (d);

            \draw [->,thick] (root) to[bend left=30] (d);
            \draw [->,thick] (X) -- (a);
            \draw [->,thick] (X) -- (b);
            \draw [->,thick] (X) -- (d);
            \draw [<->,thick] (X) to[bend right=70] (c);
        \end{tikzpicture}
        \caption{Checking $d \to^+ c$. All nodes remain reachable, so there is
        a path $d \to^+ c$.}
        \label{fig:lb1C}
    \end{subfigure}
    \hfill
    \begin{subfigure}{0.45\textwidth}
        \center
        \begin{tikzpicture}[scale=0.8]
            \node[gcnode] (root) at (-3, 1.5) {root};
            \node[gcnode,red] (X) at (-1.5, 1.5) {X};

            \node[gcnode,red] (a) at (0, 0) {a};
            \node[gcnode,red] (b) at (1.5, 0) {b};
            \node[gcnode] (c) at (3, 0) {c};
            \node[gcnode,red] (d) at (1.5, 1.5) {d};

            \draw[thick] (a) edge[->] (b) (b) edge[->] (c);
            \draw[thick] (b) edge[<->] (d);

            \draw [->,thick] (root) to[bend right=50] (c);
            \draw [<->,thick] (X) -- (a);
            \draw [->,thick] (X) -- (b);
            \draw [->,thick] (X) -- (d);
            \draw [->,thick] (X) to[bend right=70] (c);
        \end{tikzpicture}
        \caption{Checking $c \to^+ a$. The node $a$ becomes unreachable, so
        there is no path $c \to^+ a$.}
        \label{fig:lb1D}
    \end{subfigure}
    \caption{Illustration of the reduction from DRDS to GCDS~(Algorithms~\ref{alg:lbinsert}--\ref{alg:lbquery}).}
    \label{fig:lb1}
\end{figure}

\begin{example}
    \pref{fig:lb1} illustrates our reduction constructing a DRDS given an GCDS.
    As sketched, the GCDS heap will look like a copy of the DRDS graph, except
    with an auxiliary node $X$ that has edges from the root and to every other
    node (compare~\pref{fig:lb1A}~and~\pref{fig:lb1B}).  DRDS insertion and
    deletion operations are passed directly to their corresponding GCDS
    operation. To check for a path $d \to^+ c$, we add edges $\GCRoot \to d$
    and $c \to X$, then $\GCRoot \to X$~(\pref{fig:lb1C}). If there is a path
    $d \to^+ c$, then $c$ will remain reachable from $\GCRoot$ and hence so
    will $X$ and hence so will every other node~(\pref{fig:lb1C}).  Otherwise,
    if there is no such path, as in~\pref{fig:lb1D} where we have checked for a
    path $c \to^+ a$, at least the node $X$ will become unreachable and so
    $\GCDelete$ will either report unreachable nodes or time out. In either
    case, we know whether the path exists or not.
    \qed
\end{example}

We can now prove the reduction correct.
\begin{theorem}
    \label{thm:GeneralConnected}
    Suppose a GCDS has all-reachable worst-case pause times
    $\tAR(n)$ and worst-case delay
    $d(n)$~(Definitions~\ref{def:ARTime}, \ref{def:Delay}).
    Then there exists a DRDS taking time $O(\tAR(n)d(n))$ per
    operation.
\end{theorem}
\begin{proof}
    \pref{fig:LBAlgs} constructs such a DRDS assuming access to such a GCDS.
    $\DCInsert$ and $\DCDelete$ call the corresponding GC operation, ensuring
    the GCDS sees a mirror of the graph.

    $\DCConnected(v_i \to^+ v_j)$ is slightly more complicated. We insert the
    edge $\GCRoot \to v_i$ and remove the edge $\GCRoot \to X$. Hence, the root
    is connected directly \emph{only} to $v_i$, and $v_j$ will be reachable
    from $\GCRoot$ if and only if there is a path $v_i \to^+ v_j$. But because
    $v_j \to X$ and $X$ has an edge to every other node, the entire heap
    remains reachable if and only if there is a path $v_i \to^+ v_j$.
    Hence, after timing out $\GCDelete$ in order to use it as a binary decider
    whether any node has become unreachable, the result tells us whether there
    is a path $v_i \to^+ v_j$.
\end{proof}
Note that the theorem holds even if we do not know the precise expression for
$\tAR(n)$ and $d(n)$; as long as we know an asymptotic upper bound, it is
\emph{possible} (in theory) to carry out the construction in~\pref{fig:LBAlgs}
and hence the lower bound holds.
We now note the following important corollaries:
\begin{corollary}
    \label{cor:LB1}
    Any correct GCDS must have $\tAR(n)d(n) = \TOmega(\log^{3/2}{n})$.
\end{corollary}
\begin{proof}
    By the prior theorem and that of~\citet{bestlbdc}.
\end{proof}
\begin{corollary}
    \label{cor:LB2}
    Any correct GCDS must have $t(n)d(n) = \TOmega(\log^{3/2}{n})$.
\end{corollary}
\begin{proof}
    By the prior corollary and~\pref{lem:Ts}.
\end{proof}

\subsection{Hints to the Data Structure}
Our reduction has the following property: if the DRDS graph is acyclic, then
all cycles in the heap seen by the GCDS go through the $X \uto v_j$ edge.
Because the lower bound from~\citet{bestlbdc} holds even when restricted to
acyclic graphs, no GCDS can beat the $\TOmega(\log^{3/2}{n})$ lower bound
\emph{even if} the data structure is told that all cycles are cut by a single
edge, and given that edge.
This is an example where weak--strong pointer annotations do not help: the edge
is necessary to reach many nodes, hence it cannot be weak, but also inherently
introduces cycles, hence it cannot be strong.

\subsection{Conditional Lower Bounds}
The graph algorithms community has proposed a problem,
the \emph{online matrix-vector multiplication problem} (OMV)~\citep{omv}, that they
conjecture to be hard.
Similar to the exponential time conjecture about SAT, the OMV conjecture
implies other problems should have certain lower bounds as well.
In particular, the OMV conjecture would imply that the DRDS problem is lower
bounded by $\Omega(n)$ and hence it would imply any GCDS has $\tAR(n)d(n) =
\Omega(n)$ and therefore also that $t(n)d(n) = \Omega(n)$.

\section{Sparse, Acyclic Lower Bound}
\label{sec:LowerBounds2}
In the last section we proved $\tAR(n)d(n) = \tilde{\Omega}(\log^{3/2}{n})$,
implying that for any GCDS with $d(n) = O(1)$ there is a sequence of operations
where (i) one of the operations takes time at least
$\tilde{\Omega}(\log^{3/2}{n})$ and (ii) none of the operations makes anything
unreachable.
However, no clear guarantee was made about the shape of the heap in that
worst-case-triggering sequence, e.g., triggering the
$\tilde{\Omega}(\log^{3/2}{n})$-time pause might require a heap with cycles or
many outgoing edges from each region.

To address these concerns we prove in this section that $\tSAAR(n)d(n) =
\Omega(\log{n})$.
In other words, for any GCDS with $O(1)$ worst-case collection delay there is a
sequence of operations (in fact, a family of sequences of operations, one for
each $n$) where:
\begin{enumerate}
    \item One of the operations takes time at least
        $\Omega(\log{n})$;
    \item None of the operations makes anything unreachable;
    \item At all points, the heap is acyclic; and
    \item At all points, the heap is sparse.
\end{enumerate}
(\pref{sec:Algorithm} will prove that the reduction in bound from $\log^{3/2}{n}$
to $\log{n}$ is unavoidable.)

\subsection{Layered Permutation Graph Reachability}
\label{sec:LPGCDS}
We are not aware of any way to prove the desired result using the result
of~\citet{bestlbdc}.
Instead, this section presents a reduction from the \emph{layered
permutation reachability} problem, defined on the next page.
\clearpage
\begin{definition}
    \label{def:LPRDS}
    A \emph{layered permutation reachability data structure} (LPRDS) is
    identical to a DRDS~(\pref{def:DRDS}) except each vertex is assigned a
    layer number and the caller promises:
    \begin{enumerate}
        \item Edges will only be inserted from one layer to the subsequent one,
        \item Reachability queries always start from the first layer, and
        \item Every vertex has indegree and outdegree at most one.
    \end{enumerate}
    To distinguish LPRDS operations from DRDS, we call them $\LPCInsert$,
    $\LPCDelete$, $\LPCConnected$.
\end{definition}

\citet{lbdc} prove that any such data structure requires at
least~$\Omega(\log{n})$ time in the cell probe
model~(\pref{sec:CPM})~\citep{cpm}, even when amortized over arbitrarily long
sequences of operations.
The presentation here differs slightly from that of~\citet{lbdc} because
they consider connectivity in undirected graphs and do not provide the
algorithm with the layer information directly; \pref{app:LPRDS} explains how
lower bounds from their work apply to this problem as well.

\begin{theorem} (\citep{lbdc})
    Every LPRDS has worst-case~$\Omega(\log{n})$ per-operation time.
\end{theorem}

\subsection{Our Reduction}
We will assume access to a GCDS that is efficient on sequences of
operations where nothing becomes unreachable and the heap remains sparse and
acyclic.
We must use this GCDS to build an efficient algorithm for the LPRDS problem.

The reduction we will use is very similar to the general one described
in~\pref{sec:LB1}.
However, that reduction relied fundamentally on the insertion of a cycle,
namely, $X \uto v_j$, so it seems unclear how to make use of the assumption
about a time bound that applies only when the heap is acyclic.
The key is to use a slightly different reduction, shown
in~Algorithms~\ref{alg:lb2insert}--\ref{alg:lb2query}. The below example
illustrates this reduction.

\begin{figure}[t]
    \begin{minipage}{0.49\textwidth}
        \begin{algorithm}[H]
            \caption{$\LPCInsert(v_i \to v_j)$} \label{alg:lb2insert}
            $v_i, v_j \gets \LPCCreate(i, j)$\;
            $\GCInsert(v_i \to v_j)$\;
            $\GCDelete(\GCRoot \to v_j)$\;
        \end{algorithm}
        \begin{algorithm}[H]
            \caption{$\LPCDelete(v_i \to v_j)$} \label{alg:lb2delete}
            $v_i, v_j \gets \LPCCreate(i, j)$\;
            $\GCInsert(\GCRoot \to v_j)$\;
            $\GCDelete(v_i \to v_j)$\;
        \end{algorithm}
    \end{minipage}
    \hfill
    \begin{minipage}{0.49\textwidth}
        \begin{algorithm}[H]
            \caption{$\LPCCreate(i_1, i_2, \ldots)$} \label{alg:lb2init}
            \For{$i_k \in i_1, i_2, \ldots$}{
                \If{$i_k \not\in N$}{
                    $N(i_k) \gets \GCAllocate()$\;
                }
            }
            \Return{$N(i_1), N(i_2), \ldots$}\;
        \end{algorithm}
    \end{minipage}
    \begin{algorithm}[H]
        \caption{$\LPCConnected(v_i \to^+ v_j)$} \label{alg:lb2query}
        \textbf{\textsc{Checkpoint}}\;
        $v_i, v_j \gets \LPCCreate(i, j)$\;
        $\GCInsert(v_j \to v_i)$\;
        \If{$\GCDelete(\GCRoot \to v_i)$ runs for more than $\tSAAR(n)$ instructions or $\GCFreeList$ is not empty}{
            \textbf{\textsc{Restore}}\;
            \Return{\textsc{Reaches}}\;
        }
        \ForEach{$i \in 1, 2, \ldots, d(n)$}{
            \If{$\GCStep()$ runs for more than $\tSAAR(n)$ instructions or $\GCFreeList$ is not empty}{
                \textbf{\textsc{Restore}}\;
                \Return{\textsc{Reaches}}\;
            }
        }
        \textbf{\textsc{Restore}}\;
        \Return{\textsc{NotReaches}}\;
    \end{algorithm}
    \caption{Reduction from LPRDS to GCDS.}
    \label{fig:LB2Algs}
\end{figure}

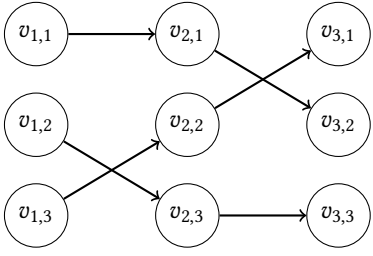
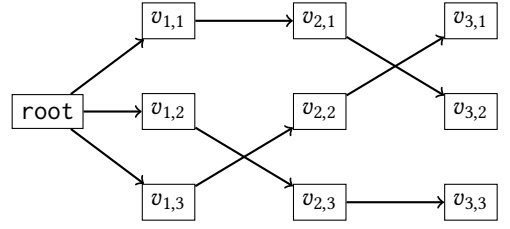
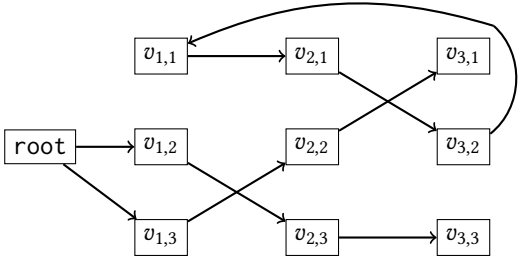
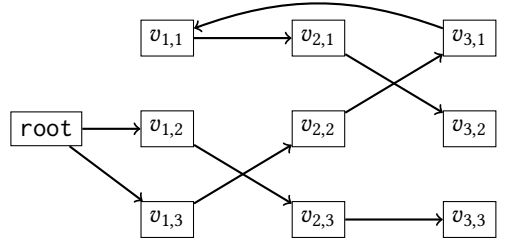
\begin{figure}[t]
    \centering
    \begin{subfigure}[t]{0.47\textwidth}
        \center
        \begin{tikzpicture}[scale=0.8]
            \node[gnode] (v11) at (0, 0) {$v_{1,1}$};
            \node[gnode] (v12) at (0, -1.5) {$v_{1,2}$};
            \node[gnode] (v13) at (0, -3) {$v_{1,3}$};
            \node[gnode] (v21) at (2.5, 0) {$v_{2,1}$};
            \node[gnode] (v22) at (2.5, -1.5) {$v_{2,2}$};
            \node[gnode] (v23) at (2.5, -3) {$v_{2,3}$};
            \node[gnode] (v31) at (5, 0) {$v_{3,1}$};
            \node[gnode] (v32) at (5, -1.5) {$v_{3,2}$};
            \node[gnode] (v33) at (5, -3) {$v_{3,3}$};

            \draw[thick] (v11) edge[->] (v21) (v21) edge[->] (v32);
            \draw[thick] (v12) edge[->] (v23) (v23) edge[->] (v33);
            \draw[thick] (v13) edge[->] (v22) (v22) edge[->] (v31);
        \end{tikzpicture}
        \caption{A layered permutation graph as in~\citet{lbdc}.}
        \label{fig:lb2A}
    \end{subfigure}
    \hfill
    \begin{subfigure}[t]{0.47\textwidth}
        \center
        \begin{tikzpicture}[scale=0.8]
            \node[gcnode] (root) at (-2, -1.5) {$\GCRoot$};
            \node[gcnode] (v11) at (0, 0) {$v_{1,1}$};
            \node[gcnode] (v12) at (0, -1.5) {$v_{1,2}$};
            \node[gcnode] (v13) at (0, -3) {$v_{1,3}$};
            \node[gcnode] (v21) at (2.5, 0) {$v_{2,1}$};
            \node[gcnode] (v22) at (2.5, -1.5) {$v_{2,2}$};
            \node[gcnode] (v23) at (2.5, -3) {$v_{2,3}$};
            \node[gcnode] (v31) at (5, 0) {$v_{3,1}$};
            \node[gcnode] (v32) at (5, -1.5) {$v_{3,2}$};
            \node[gcnode] (v33) at (5, -3) {$v_{3,3}$};

            \draw[thick] (root) edge[->] (v11) edge[->] (v12) edge[->] (v13);
            \draw[thick] (v11) edge[->] (v21) (v21) edge[->] (v32);
            \draw[thick] (v12) edge[->] (v23) (v23) edge[->] (v33);
            \draw[thick] (v13) edge[->] (v22) (v22) edge[->] (v31);
        \end{tikzpicture}
        \caption{The corresponding heap graph we use for the lower bound
        reduction.}
        \label{fig:lb2B}
    \end{subfigure}
    \begin{subfigure}{0.45\textwidth}
        \center
        \begin{tikzpicture}[scale=0.8]
            \node[gcnode] (root) at (-2, -1.5) {$\GCRoot$};
            \node[gcnode] (v11) at (0, 0) {$v_{1,1}$};
            \node[gcnode] (v12) at (0, -1.5) {$v_{1,2}$};
            \node[gcnode] (v13) at (0, -3) {$v_{1,3}$};
            \node[gcnode] (v21) at (2.5, 0) {$v_{2,1}$};
            \node[gcnode] (v22) at (2.5, -1.5) {$v_{2,2}$};
            \node[gcnode] (v23) at (2.5, -3) {$v_{2,3}$};
            \node[gcnode] (v31) at (5, 0) {$v_{3,1}$};
            \node[gcnode] (v32) at (5, -1.5) {$v_{3,2}$};
            \node[gcnode] (v33) at (5, -3) {$v_{3,3}$};

            \draw[thick] (root) edge[->] (v12) edge[->] (v13);
            \draw[thick] (v11) edge[->] (v21) (v21) edge[->] (v32);
            \draw[thick] (v12) edge[->] (v23) (v23) edge[->] (v33);
            \draw[thick] (v13) edge[->] (v22) (v22) edge[->] (v31);
            \draw[thick] (v32) edge[bend right=50] (5.5, 0.5) (5.5, 0.5) edge[->,bend right=20] (v11);
        \end{tikzpicture}
        \caption{Checking $v_{1,1} \to^+ v_{3,2}$. A cycle is introduced and
        $v_{1,1}$ becomes unreachable, so there is a path $v_{1,1} \to^+ v_{3,2}$.}
        \label{fig:lb2C}
    \end{subfigure}
    \hfill
    \begin{subfigure}{0.45\textwidth}
        \center
        \hspace{-4mm}
        \begin{tikzpicture}[scale=0.8]
            \node[gcnode] (root) at (-2, -1.5) {$\GCRoot$};
            \node[gcnode] (v11) at (0, 0) {$v_{1,1}$};
            \node[gcnode] (v12) at (0, -1.5) {$v_{1,2}$};
            \node[gcnode] (v13) at (0, -3) {$v_{1,3}$};
            \node[gcnode] (v21) at (2.5, 0) {$v_{2,1}$};
            \node[gcnode] (v22) at (2.5, -1.5) {$v_{2,2}$};
            \node[gcnode] (v23) at (2.5, -3) {$v_{2,3}$};
            \node[gcnode] (v31) at (5, 0) {$v_{3,1}$};
            \node[gcnode] (v32) at (5, -1.5) {$v_{3,2}$};
            \node[gcnode] (v33) at (5, -3) {$v_{3,3}$};

            \draw[thick] (root) edge[->] (v12) edge[->] (v13);
            \draw[thick] (v11) edge[->] (v21) (v21) edge[->] (v32);
            \draw[thick] (v12) edge[->] (v23) (v23) edge[->] (v33);
            \draw[thick] (v13) edge[->] (v22) (v22) edge[->] (v31);
            \draw[thick] (v31) edge[->,bend right=20] (v11);
        \end{tikzpicture}
        \caption{Checking $v_{1, 1} \to^+ v_{3, 1}$. The graph remains acyclic
        and nothing becomes unreachable, so there is a path $v_{1, 1} \to^+ v_{3, 1}$.}
        \label{fig:lb2D}
    \end{subfigure}
    \caption{Illustration of the reduction from the problem of~\citet{lbdc} to
    GCDS.}
    \label{fig:lb2}
\end{figure}

\begin{example}
    Consider the layered permutation graph in~\pref{fig:lb2A}.
    Our goal is to use an efficient GCDS to quickly answer reachability
    queries in such graphs.
    Our reduction makes a copy of the layered permutation graph in the GCDS,
    and connects $\GCRoot$ to every node with no incoming edges so the graph
    remains connected~(\pref{fig:lb2B}).
    To check for a path~$a \to^+ b$, we add an edge from $b \to a$ and then
    disconnect $a$ from $\GCRoot$.
    If $a$ \emph{can} reach $b$, i.e., they are on the same
    path~(\pref{fig:lb2C}), then a cycle will be inserted and both will become
    unreachable.
    If $a$ \emph{cannot} reach $b$, i.e., they belong to different
    paths, then $b$ will remain reachable and hence so will $a$ and everything
    else along its path without the introduction of any
    cycles~(\pref{fig:lb2D}).
    In either case, the presence of cycles or unreachable nodes tells us
    whether a path exists.
    Because we are assuming the GCDS is fast when no cycles are present and
    nothing becomes unreachable, we can use the same timeout idea to quickly
    check for these conditions.
    \qed
\end{example}

We can now prove the main theorem of this section.

\begin{theorem}
    \label{thm:AcyclicLB}
    Suppose some GCDS has sparse, acyclic, all-reachable worst-case pause times
    $\tSAAR(n)$ and worst-case delay $d(n)$.
    Then there exists an LPRDS taking time $O(\tSAAR(n)d(n))$ per operation.
\end{theorem}

\begin{proof}
    The algorithms for the reduction are given in~\pref{fig:LB2Algs}.
    We mirror the graph into the GCDS heap, but add an edge from $\GCRoot$ to
    every node with indegree zero.
    This ensures every node is reachable via a unique path from $\GCRoot$.

    To check if $v_i \to^+ v_j$, i.e., whether $v_j$ is in the path starting at
    $v_i$, we add $v_j \to v_i$ then disconnect $v_i$ from $\GCRoot$.
    If $v_j$ \emph{is} reachable from $v_i$, we introduced a cycle and severed
    the unique path to $v_i$ and $v_j$.
    Otherwise, no cycle is inserted and $v_i$ remains accessible via the path
    from $v_j$.
    In either case, the algorithm correctly reports the (non)existence of such
    a path to the caller.
\end{proof}

Hence, we immediately get the following two corollaries:
\begin{corollary}
    \label{cor:SAARLB1}
    Any correct GCDS must have $\tSAAR(n)d(n) = \Omega(\log{n})$.
\end{corollary}
\begin{proof}
    By the prior theorem and that of~\citet{bestlbdc}.
\end{proof}

\subsection{Connection to Cyclic Reference Counting}
\label{sec:Sanity}
The results in this section connect closely to the field of \emph{cyclic
reference counting}, which is focused on extending reference counting to
support heaps with cycles.
One of the first algorithms, from~\citet{brownbridge85}, claimed to support
immediate collection of all unreachable regions, even in the presence of
cycles, while ensuring $O(1)$ pause times for acyclic heaps, i.e.,
violating~\pref{cor:SAARLB1}.
But it was discovered to be incorrect by~\citet{salkild}.
If the lower bounds in this section were known at the time, sanity checking
that something is wrong with the claims would have been significantly easier.
More recently, the English translation of~\citet{ejh} also claims to beat our
lower bound, i.e., guarantee immediate collection for all heaps and also
guarantee $O(1)$ pause times when the heap is acyclic.
We believe this claim is simply a mistranslation; nonetheless, we provide
in~\pref{app:PepelsCex} a counterexample to the claim.

\section{Application-Level Implications of Lower Bounds}
\label{sec:SummaryLB}
We now briefly review the application-level consequences of the lower bounds in
the last two sections.
Consider a programming language with a garbage collector implementing the
relatively standard GCDS interface, i.e., in an abstract sense, it sees the
program as a sequence of pointer insertions and removals.

\subsection{Main Lower Bound, Existence of Real-Time Violations}
Suppose the language enforces a memory limit $M$.
Suppose the underlying collector enforces an $O(1)$ pause time limit, i.e.,
$t(n) = O(1)$.
\pref{thm:GeneralConnected} and \pref{cor:LB1} imply there exists at least one
program of the following form:
\begin{enumerate}
    \item First fill the memory up to the limit with equal-sized allocations
        and perform some pointer manipulations while keeping every region
        reachable;
    \item Then, overwrite a pointer to make some region unreachable (collectable);
    \item Then, request a new allocation;
\end{enumerate}
such that the region made collectable in the second step takes
$\TOmega(\log^{3/2}{M})$ program steps to collect.
Hence, in the third step, either the program must introduce a
$\TOmega(\log^{3/2}{M})$-time pause to complete collection, or it must
incorrectly report out-of-memory.
Either case could be catastrophic in real-time contexts.

\subsection{Acyclic Lower Bound}
Suppose again that the language attempts to guarantee for the user that
finalizers are called reliably and promptly, i.e., $d(n) = O(1)$.
The second lower bound also implies that there must exist a program, even a
program where nothing ever becomes unreachable \emph{and the heap stays sparse
and acyclic}, where this language introduces an $\Omega(\log{n})$-length pause
time after some program operation.
Equivalently, reference counting can not be extended to handle cycles without
slowing down execution on some program that only makes acyclic heap structures.

\section{Upper Bound}
\label{sec:Algorithm}
\pref{sec:LowerBounds2} showed that any GCDS guaranteeing constant collection
delay for all programs ($d(n) = O(1)$) must introduce significant slowdown on
some program making only a sparse, acyclic heap, even if that program never
makes any region unreachable ($\tSAAR(n) = \Omega(\log{n})$).

This section shows that this bound is tight: there is a GCDS that guarantees
constant collection delay for all programs ($d(n) = 1$), and guarantees that no
operation takes longer than $O(\log{n})$ time when the heap is acyclic and
nothing becomes unreachable ($\tAAR(n) = O(\log{n})$).
In fact, we will eventually show the stronger fact that, when the heap is
acyclic, our GCDS has exactly an $O(\log{n})$-factor slowdown compared to
reference counting.

The key difference with reference counting, however, is that our algorithm
still guarantees constant collection delay even in the presence of cycles.
Combined with the lower bound in the prior section, this algorithm has optimal
acyclic, all-reachable worst-case pause times among all algorithms that
guarantee $d(n) = O(1)$.
However, because it involves both frequent balanced tree operations and
increased (though still constant sized) metadata per allocation region, we have
found that it is not competitive in non-pathological cases with traditional
garbage collection algorithms.
Hence, we do not recommend it as a general-purpose GC; its primary purpose is
to answer the theoretical questions described above.

The algorithm itself is adapted from similar algorithms for dynamic
connectivity in undirected graphs~\citep{leveltrees,xortrick}.
Like those, the key idea is to maintain a spanning tree of the reachable heap
in an Euler tour data structure~(\pref{sec:ET}).
The main complication is updating the spanning tree when an edge in the
spanning tree is removed from the heap.
Prior work has devised many clever techniques for quickly updating the spanning
tree when the graph is undirected, however, these techniques are not
immediately applicable to the directed case needed for proper GC.
Instead, we identify a strategy for updating spanning trees of directed graphs
that is sufficient to guarantee the $\O{\log{n}}$ time when the graph is
acyclic.

\subsection{Euler Tour Data Structure}
\label{sec:ET}
Our algorithm relies on the Euler tour data structure, which is a well-known
data structure for efficiently storing and manipulating directed forests.
A \emph{directed tree} is a graph $(V, E)$ that has exactly $\abs{V} - 1$ edges
and a distinguished \emph{root} vertex that can reach every other node.
A \emph{directed forest} is a disjoint union of directed trees.
The Euler Tour data structure stores a directed forest while allowing efficient
edge insertions, edge deletions, and connectedness queries.
\begin{definition}
    The Euler Tour data structure (ETDS) stores directed forests and supports:
    \begin{itemize}
        \item $\ETSingleton()$ returns a new singleton tree in the forest.
        \item $\ETCut(a \to b)$ splits a tree into two by removing the edge $a \to
            b$.
        \item $\ETJoin(a \to b)$ links a tree containing $a$ and one rooted at
            $b$ by inserting the edge $a \to b$.
        \item $\ETPath(a \to^+ b)$ is true if and only if there is a directed
            path from $a$ to $b$ in the forest.
        \item $\ETParent(a)$ returns the immediate parent of $a$ in the tree.
        \item $\ETNext(a)$ returns the node after $a$ in the preorder traversal
            of the tree.
    \end{itemize}
\end{definition}
There exist deterministic algorithms for the Euler tour data structure where
$\ETSingleton$, $\ETCut$, $\ETJoin$, and $\ETPath$ all take time $O(\log{n})$
while $\ETNext$ and $\ETParent$ take time $O(1)$.
The insight is to represent each tree by its \emph{Euler tour}.
This Euler tour is itself stored in a balanced binary tree.
See~\citet{ett} for a detailed discussion of the algorithms.

\subsection{Algorithm and Pseudocode}
\begin{figure}
    \begin{minipage}[t]{0.48\textwidth}
        \begin{algorithm}[H]
            \caption{$\GCInit()$} \label{alg:gcinit}
            $\GCRoot \gets \ETSingleton()$\;
        \end{algorithm}
        \begin{algorithm}[H]
            \caption{$\GCInsert(a \to b)$} \label{alg:gcinsert}
            $E(a \to b) \gets E(a \to b) + 1$\;
        \end{algorithm}
    \end{minipage}
    \hfill
    \begin{minipage}[t]{0.48\textwidth}
        \begin{algorithm}[H]
            \caption{$\GCAllocate()$} \label{alg:gcallocate}
            $n \gets \ETSingleton()$\;
            $\ETJoin(\GCRoot \to n)$\;
            $E(\GCRoot \to n) \gets 1$\;
            \Return{$n$}\;
        \end{algorithm}
    \end{minipage}
\begin{algorithm}[H]
    \caption{$\GCStep()$} \label{alg:gcstep}
    \textbf{no-op}\;
\end{algorithm}
\begin{algorithm}[H]
    \caption{$\GCDelete(a \to b)$} \label{alg:gcdelete}
    $E(a \to b) \gets E(a \to b) - 1$\;
    \lIf{$E(a \to b) > 0$ or $\ETParent(b) \neq a$}{
        \Return{}
    }\label{li:sameparent}
    $\ETCut(a \to b)$\;
    \While{$\top$}{
        $D \gets \emptyset$ \tcc*{Nodes earlier in the preorder traversal}
        $c \gets \bot$ \tcc*{Continue iterations, or reached fixedpoint?}
        \tcc{Iterate over the tree rooted at $b$; keep two pointers so we can
        continue the iteration even if a subtree is cut out.}
        $p, n \gets \bot, b$\;
        \While{$n$ is not none}{
            \tcc{Look for a new parent either in the main spanning tree or
            later in the preorder traversal.}
            Find $m$ such that $E(m \to n) > 0$, $m \not\in D$, and $\neg\ETPath(n \to^+ m)$\;
            \eIf{there is an $m$ satisfying those conditions}{
                $\ETCut(n)$ \tcc*{Cut subtree away from the separated spanning tree}
                $\ETJoin(m \to n)$ \tcc*{Reconnect it to the new parent}
                \lIf{$n = b$}{\Return{}}
                \lIf{$\ETPath(\GCRoot \to^+ n)$}{$c = \top$}
                $p, n \gets p, \ETNext(p)$ \tcc*{Skip that subtree}
            }{
                $D \gets D \cup \{ n \}$ \tcc*{Do not link later nodes here}
                $p, n \gets n, \ETNext(n)$ \tcc*{Continue the preorder traversal}
            }
        }
        \If{$c = \bot$}{
            $\GCFreeList{} \gets \GCFreeList{} \cup D$\;
            \Return{}
        }
    }
\end{algorithm}
\end{figure}

For ease of exposition the majority of the section describes a simplified
algorithm that only guarantees short worst-case pause times when the heap is
both sparse and acyclic.
Then in~\pref{sec:Dense} we describe minor optimizations that permit the same
guarantees when the heap is dense.
The proposed GCDS is described by~\pref{alg:gcallocate}--\ref{alg:gcdelete}.
It stores two pieces of information:
\begin{enumerate}
    \item An Euler Tour Tree representation of a spanning forest for the heap
        graph, and
    \item A map from edges to multiplicity that can be queried for all edges
        ending in a given node.
\end{enumerate}

To allocate a new node, we create the node in the ETT, then link that ETT
node to the one for $\GCRoot$.
Inserting an edge does not need to update the spanning tree because we know the
node was already reachable from $\GCRoot$ and hence already part of the
spanning tree.

Deleting an edge $a \to b$ is more difficult.
If $a$ is \emph{not} the parent of $b$, i.e., $a\to b$ is not in the spanning
tree, we simply remove it from the edges map and leave the tree unchanged.
Otherwise, if $a \to b$ was an edge in the spanning tree, we cut $b$ from the
spanning tree and begin a preorder iteration of the newly separated tree rooted
at $b$.
For each node $n$, we iterate over its incoming edges $m \to n$ looking for one
where $m$ is not a descendant of $n$ and either (i) $m$ is part of the main
spanning tree, or (ii) $m$ is in the separated tree but not yet visited.
If an incoming edge $m \to n$ satisfying those two conditions is found, we cut the
subtree rooted at $n$ out of the tree and relink it at $m$.
Intuitively, this process pushes subtrees either back into the main spanning
tree or further down the preorder traversal of the newly separated tree.
We repeat this sweeping operation until the entire sweep reveals no nodes with
incoming edges from the main spanning tree, at which point we know the entire
separated tree is unreachable.
(Note $\ETNext(p) = \bot$ only when $n=b$, hence line 15 will never be reached
with $\ETNext(p) = \bot$ --- it would instead return on line 13.)

\begin{figure}[t]
    \centering
    \begin{subfigure}[t]{0.48\textwidth}
        \center
        \begin{tikzpicture}[scale=0.8]
            \node[gcnode] (R) at (0.5, 0) {$\GCRoot$};
            \node[gcnode] (b) at (-2, 0) {$b$};
            \node[gcnode] (n1) at (-2, -1.5) {$n_1$};
            \node[gcnode] (n2) at (-1, -1.5) {$n_2$};
            \node[gcnode] (n3) at (0.5, -1.5) {$n_3$};
            \node[gcnode] (n4) at (-1, -3) {$n_4$};
            \draw[thick] (R) edge[->,densely dotted,red] (b) (b) edge[->] (n1) (b) edge[->] (n3);
            \draw[thick] (b) edge[->] (n2) (n2) edge[->,bend left=30] (n4) (n4) edge[->,bend left=30,dashed] (n2);
            \draw[thick] (R) edge[->,dashed] (n3) (n3) edge[->,dashed] (n2);
            \draw[blue] (-1.5, 0.5) rectangle (-2.5, -2);
            \node at (-2.75, -2.25) {$D$};
        \end{tikzpicture}
        \caption{Here, the algorithm already visited $b$ and $n_1$. Neither
        had an incoming edge from a node not in $D$ or from the main spanning
        tree (which is now just $\GCRoot$), so they both are added to $D$.}
        \label{fig:upperA}
    \end{subfigure}
    \hfill
    \begin{subfigure}[t]{0.48\textwidth}
        \center
        \begin{tikzpicture}[scale=0.8]
            \node[gcnode] (R) at (0.5, 0) {$\GCRoot$};
            \node[gcnode] (b) at (-2, 0) {$b$};
            \node[gcnode] (n1) at (-2, -1.5) {$n_1$};
            \node[gcnode] (n2) at (-1, -1.5) {$n_2$};
            \node[gcnode] (n3) at (0.5, -1.5) {$n_3$};
            \node[gcnode] (n4) at (-1, -3) {$n_4$};
            \draw[thick] (R) edge[->,densely dotted,red] (b) (b) edge[->] (n1) (b) edge[->] (n3);
            \draw[thick] (b) edge[->,dashed] (n2) (n2) edge[->,bend left=30] (n4) (n4) edge[->,bend left=30,dashed] (n2);
            \draw[thick] (R) edge[->,dashed] (n3) (n3) edge[->] (n2);
            \draw[blue] (-1.5, 0.5) rectangle (-2.5, -2);
            \node at (-2.75, -2.25) {$D$};
        \end{tikzpicture}
        \caption{The only edge to $n_2$ that is from a node neither in $D$ nor
        a tree-descendant of $n_2$ is $n_3 \to n_2$.  Hence, we update the
        spanning tree by cutting $n_2$ from its parent and linking it to
        $n_3$.}
        \label{fig:upperB}
    \end{subfigure}
    \hfill
    \begin{subfigure}[t]{0.48\textwidth}
        \center
        \begin{tikzpicture}[scale=0.8]
            \node[gcnode] (R) at (0.5, 0) {$\GCRoot$};
            \node[gcnode] (b) at (-2, 0) {$b$};
            \node[gcnode] (n1) at (-2, -1.5) {$n_1$};
            \node[gcnode] (n2) at (-1, -1.5) {$n_2$};
            \node[gcnode] (n3) at (0.5, -1.5) {$n_3$};
            \node[gcnode] (n4) at (-1, -3) {$n_4$};
            \draw[thick] (R) edge[->,densely dotted,red] (b) (b) edge[->] (n1) (b) edge[->,dashed] (n3);
            \draw[thick] (b) edge[->,dashed] (n2) (n2) edge[->,bend left=30] (n4) (n4) edge[->,bend left=30,dashed] (n2);
            \draw[thick] (R) edge[->] (n3) (n3) edge[->] (n2);
            \draw[blue] (-1.5, 0.5) rectangle (-2.5, -2);
            \node at (-2.75, -2.25) {$D$};
        \end{tikzpicture}
        \caption{A similar analysis finds that $n_3$ can be rooted at $\GCRoot
        \to n_3$, so the spanning tree is updated. This links $n_3$ and its
        entire subtree, including $n_2$ and $n_4$, to the main spanning tree.}
        \label{fig:upperC}
    \end{subfigure}
    \hfill
    \begin{subfigure}[t]{0.48\textwidth}
        \center
        \begin{tikzpicture}[scale=0.8]
            \node[gcnode] (R) at (0.5, 0) {$\GCRoot$};
            \node[gcnode] (n2) at (-1, -1.5) {$n_2$};
            \node[gcnode] (n3) at (0.5, -1.5) {$n_3$};
            \node[gcnode] (n4) at (-1, -3) {$n_4$};
            \draw[thick] (n2) edge[->,bend left=30] (n4) (n4) edge[->,bend left=30,dashed] (n2);
            \draw[thick] (R) edge[->] (n3) (n3) edge[->] (n2);
        \end{tikzpicture}
        \caption{A final iteration of the loop finds that nothing else can be
        connected to the main spanning tree, hence~$D=\{b, n_1\}$ can be
        deleted from the graph and freed in the program.}
        \label{fig:upperD}
    \end{subfigure}
    \caption{Visualizing the operation of $\GCDelete(\GCRoot \to b)$. The red
    dotted edge is the deleted edge. The solid edges represent edges in the
    spanning trees. The dashed edges represent edges in the heap graph that are
    not currently part of the spanning tree. The visualization starts after two
    nodes in the preorder have already been visited. We then visit $n_2$,
    relinking it to $n_3$. After visiting $n_3$ and relinking it to $\GCRoot$,
    the maximal spanning tree is restored and the remaining nodes can be
    freed. The blue box represents the set $D$ from~\pref{alg:gcdelete}.}
    \label{fig:upper}
\end{figure}
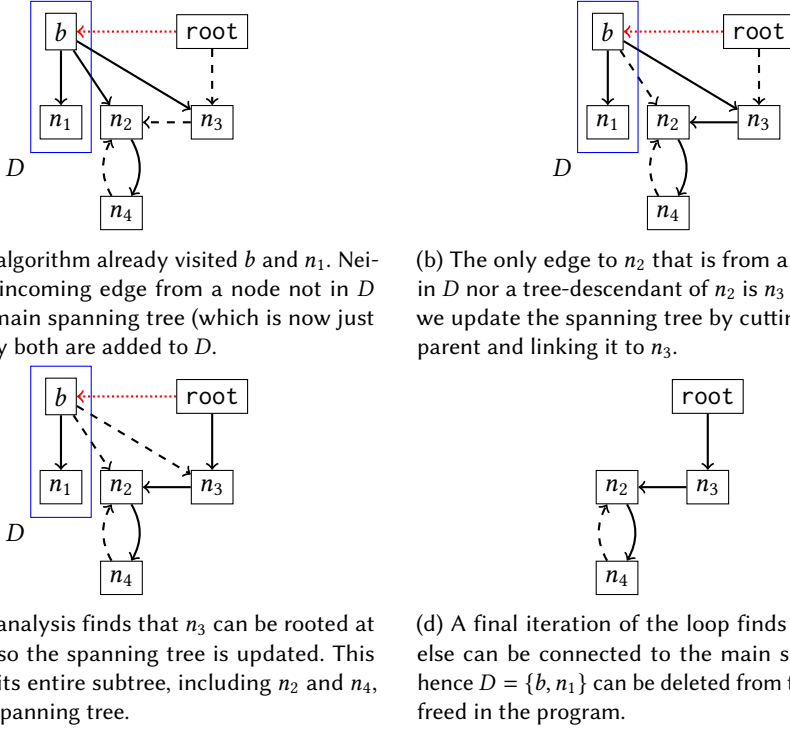

\begin{example}
    The execution of $\GCDelete$ is visualized in~\pref{fig:upper}. We
    start with a heap having six nodes, where the spanning tree is made up of
    edges $\GCRoot \to b$, $b \to n_1$, $b \to n_2$, $b \to n_3$, and $n_2 \to
    n_4$. Then, the operation $\GCDelete(\GCRoot \to b)$ is performed.

    First, the algorithm visits $b$, finding that, after removing the edge
    $\GCRoot \to b$, it has no other incoming edges at all. Hence, it is added
    to the set $D$ and we continue the preorder sweep to its child $n_1$.
    The only incoming edge to $n_1$ is $b \to n_1$, but $b \in D$ so we also
    add $n_1$ to $D$ and continue the sweep. The resulting state is
    illustrated in~\pref{fig:upperA}.

    Next in the preorder is node $n_2$. The incoming edge $b \to n_2$ is not
    usable because $b \in D$. Meanwhile, the edge $n_4 \to n_2$ is not usable
    because $n_4$ is a descendant of $n_2$ in the spanning tree --- adding that
    edge would cause a cycle. Finally, we find that the edge $n_3 \to n_2$
    works, so we unlink~$n_2$ from~$b$ and reconnect it to the spanning tree
    via~$n_3$. The resulting state is illustrated in~\pref{fig:upperB}.

    Finally we visit $n_3$. The edge from $b$ is unusable because $b \in D$,
    but the edge from $\GCRoot$ can be used. This reconnects $n_3$ to the main
    spanning tree, which now contains $n_3$, $n_2$, and $n_4$. The resulting
    state is illustrated in~\pref{fig:upperC}.

    Because we have reconnected something to the main spanning tree during this
    sweep, we must resweep across the remaining two nodes in the
    spanning tree rooted at $b$ ($b$ and $n_1$). Neither
    can be reconnected to the main spanning tree, i.e., the one rooted at
    $\GCRoot$, so the algorithm completes by removing those nodes and adding
    them to $\GCFreeList$. \qed
\end{example}

With the operation of the algorithm now illustrated, we can prove correctness
and termination.
\begin{theorem}
    \label{thm:AlgCorrect}
    Algorithms~\ref{alg:gcinit}--\ref{alg:gcdelete} terminate and correctly
    implement a GCDS with $d(n) = 1$.
\end{theorem}
\begin{proof}
    The algorithm maintains a maximal spanning tree from $\GCRoot$.  This
    property follows immediately for $\GCInit()$, $\GCInsert()$, and
    $\GCAllocate()$.
    To see correctness of $\GCDelete$, first note we only ever call $\ETLink$
    on edges that actually exist in the graph, so the spanning tree is at least
    a subset of the maximal spanning tree.
    Second, note that a node is only added to $D$ if it has no incoming edges
    from the main spanning tree, and the algorithm only halts when $D$ contains
    all the nodes in the newly separated tree.
    Hence, there can be no edges from the main spanning tree to any node in the
    newly separated tree, i.e., the main spanning tree is indeed maximal.

    For termination, it suffices to note that on every iteration of the outer
    loop except the last at least one node is moved to the main spanning tree.
    To see that $d(n) = 1$, it suffices to note that $\GCDelete$ eagerly adds
    any newly unreachable nodes to the free list.
\end{proof}

\subsection{Sparse, Acyclic, All-Reachable Worst-Case Pause Time}
We now prove that the GCDS has $\tSAAR(n) = O(\log{n})$, i.e., worst-case
logarithmic pause times when the heap is sparse and acyclic and nothing becomes
unreachable.
The time bounds for all operations except for $\GCDelete$ follow immediately
from those of the underlying data structures, so we primarily focus on
$\GCDelete$.

We can prove the stronger fact that, when the heap is sparse and acyclic,
$\GCDelete$ runs in time a logarithmic factor slower than the number of newly
unreachable edges.
Because reference counting runs in time proportional to the number of newly
unreachable edges, this implies our GCDS runs with only a logarithmic factor
overhead compared to eager reference counting as long as the heap is sparse and
acyclic.
We will use the symbol $\Delta$ for the number of edges that become
unreachable; this is identical to the number of counter decrements that
reference counting would have to do.
\begin{theorem}
    \label{thm:UpperBound}
    If $b$ cannot reach any cycles, every node reachable from $b$ has $O(1)$
    indegree, and $\Delta$ edges become unreachable after removing $a \to b$,
    then $\GCDelete(a \to b)$ runs in time $\O{\Delta \log{n}}$.
\end{theorem}
\begin{proof}
    In the acyclic case, $\ETPath(n \to^+ m)$ is impossible. Hence, a node is
    only added to $D$ when all of its predecessors are in $D$. Because $D$
    is initially empty, induction ensures~$D$ contains only nodes that are
    now unreachable from~$\GCRoot$.
    This implies the outer loop iterates exactly twice, as the first
    iteration identifies and links all still-reachable nodes back to the main
    spanning tree.
 
    Because we skip over subtrees when a new parent is found, we only visit a
    node when its parent was added to $D$. In that case, the parent is
    unreachable so the edge from the parent to the node is counted by $\Delta$.
    This ensures we do not visit any node too many times.
    Finally, thanks to sparsity, only $O(1)$ incoming edges must be analyzed
    for each visited node. Each such iteration costs $O(\log{n})$ to check for
    the existence of a path.
    Together, these imply the desired $O(\Delta \log{n})$ time complexity.
\end{proof}

We now note the following corollary:
\begin{corollary}
    The GCDS guarantees $d(n) = 1$ and $\tSAAR(n) = O(\log{n})$.
\end{corollary}
\begin{proof}
    The delay claim follows from~\pref{thm:AlgCorrect}. The time follows
    from~\pref{thm:UpperBound} ($\tSAAR(n)$ only counts operations where
    nothing becomes unreachable, i.e., $\Delta = 1$ in~\pref{thm:UpperBound}).
\end{proof}

\subsection{Handling Dense Heaps}
\label{sec:Dense}
The above time bound is within an $O(\log{n})$ factor of reference counting when
the graph is sparse and acyclic.
The sparsity requirement can be eliminated by memoizing edge information when
searching for a new parent.
If the same node is visited twice along a single sweep, any incoming edges that
were invalid during the first time it was visited are still invalid the second
time.
Hence, we can keep a pointer for each node to the last edge we examined,
avoiding revisiting edges in later iterations.
This guarantees, in the acyclic case, that you only look at
truly dead edges, i.e., at most $\O{\Delta}$ edges, for an overall time
$\O{\Delta\log{n}}$, a log-factor overhead compared to reference counting.

\subsection{Memory Overhead}
When a node is added to the free list, it will never be referred to again.
Hence, $\GCDelete$ can free any related internal data structure memory, e.g.,
the corresponding Euler tour tree nodes.
The overall scheme then has constant-factor memory overhead: every nonfreed
memory region has a corresponding ETT node, and every heap pointer has a
corresponding entry in the edge map.

\subsection{Exploratory Implementation and Evaluation}
\label{sec:Eval}
This paper is primarily theoretical, but we wondered how the above algorithm
would perform in practice.
We implemented a variant of it as a GC for Lua.
Lua was a convenient choice for implementation as it is a popular language with
a relatively simple implementation.
It uses a tricolor tracing GC, which is considerably more involved than the
na\"ive eager mark-and-sweep approach but significantly less sophisticated than
modern GCs used by Java, Javascript, etc.

We also compared to an existing GC algorithm that can be configured to
guarantees $O(1)$ collection delay, namely, we translated Figure 2
(``Synchronous Cycle Collection,'' SynCC) of~\citep{bacon_crc} into C and ran
its \texttt{CollectCycles} routine after every pointer decrement.
Notably, this always-collect configuration is not suggested
by~\citep{bacon_crc}, but it does guarantee immediate collection in the
presence of cycles (the same guarantee our ETT-based algorithm makes).
Experiments were performed on an Intel(R) Core(TM) i9-13900, with 32 GB of
memory running Debian 12 and the \texttt{jemalloc} allocator.
Other evaluation details are in~\pref{app:ExperimentDetails}.

\subsubsection{Motivating GC Thrashing Example}
Recall the motivating example from~\pref{sec:MotivateThrashing}, where frequent
collections were needed in a memory-constrained scenario.
The resulting GC thrashing increased the time from approximately 1.1 seconds
for a version of the program using manual memory management to almost 70
seconds when the limit was enforced.
We reran the same benchmark with both SynCC and our proposed GC.
SynCC took approximately 1.1 seconds while ours took 1.4 seconds.
SynCC performs better because the heap is simple and does not have any long
pointer chains; in~\pref{sec:EvalGeneral} we will see scenarios where we beat
SynCC by multiple orders of magnitude.
This example clearly demonstrates a need for fast, prompt collection and shows
that GCs with constant delay can sometimes improve performance over a
traditional tracing collector.

\subsubsection{General Benchmarks}
\label{sec:EvalGeneral}
Finally, \pref{tab:EvalGeneral} shows the timing results for multiple Lua
benchmarks, most taken from~\citep{plb}.
The default Lua collector usually has better end-to-end time than both of the
immediate collectors.
This is unsurprising, because it makes no guarantees about when or whether
regions will be collected and finalizers will be called.
In fact, the Lua interpreter is partially optimized for implementation size and
simplicity so its GC is na\"ive vs.\ those in Java, Javascript, etc.
Hence, we would expect on such languages there would be an even wider gap
between the default collector's performance and the two immediate collectors.

Between the two collectors guaranteeing $d(n) = 1$, our time overhead is
consistently better than that of SynCC.
This is highlighted by the \texttt{list-4096} benchmark, which builds and then
traverses a long linked list (the \texttt{dbllist-4096} is similar, except uses
a doubly linked list).
Our technique guarantees an asymptotic $O(\log{n})$ factor overhead for such
acyclic structures.
On the other hand SynCC must scan nearly the entire heap after every operation on
a node, introducing an asymptotic $O(n)$ factor overhead on program
operations.
Notably, our approach has a comparably high memory overhead due to the need to
store metadata (splay tree nodes for the ETDS).
Although this memory overhead is a constant factor of the number of
allocations, it is larger than that needed for reference counting or
nonimmediate collection.
Hence we do not suggest our algorithm as a good general-purpose collector, but
rather an interesting and nonobvious point in the space of asymptotic
tradeoffs that could perhaps inspire future applications.

\begin{table}[t]
    \caption{Times are averaged over 5 runs.
    Overhead is reported as a ratio compared to the default Lua interpreter;
    $1\times$ means no overhead.
    ``Nonimmediate'' is the default Lua collector, while ``SynCC'' is another
    collector also guaranteeing $d(n) = 1$ (see description
    in~\pref{sec:Eval}).
    Guaranteeing prompt collection usually takes additional time, but our
    approach was faster than SynCC.
    Our approach has higher memory overhead due to additional metadata overhead
    per allocation region to store the ETDS.
    The reader should be careful to note that the default Lua collector
    (``Nonimmediate'') is relatively na\"ive compared to the standard Java,
    JavaScript, etc., collectors, hence in corresponding experiments
    for those languages we would expect the ``Nonimmediate'' times and memory
    usage to be lower and the overhead columns to be higher.
    }
    \small
    \begin{tabular}{lrrrrrr}
        \toprule
        & \multicolumn{2}{c}{Nonimmediate} & \multicolumn{2}{c}{SynCC Overhead} & \multicolumn{2}{c}{Our Overhead} \\
        \cmidrule{2-3} \cmidrule{4-5} \cmidrule{6-7}
        Benchmark                    & Time (s) & Mem (B)& Time           & Mem            & Time         & Mem          \\
        \midrule
        \texttt{binarytrees-15}    & 0.3692             & 13044566.00       & $3.1642\times$       & $1.2950\times$    & $2.1207\times$    & $2.9869\times$    \\
        \texttt{helloworld}        & 0.0016             & 97413.00          & $0.9957\times$       & $1.1572\times$    & $0.9874\times$    & $1.6261\times$    \\
        \texttt{merkletrees-15}    & 0.9468             & 20163331.00       & $3.3933\times$       & $1.6099\times$    & $1.7635\times$    & $3.5153\times$    \\
        \texttt{nbody-100000}      & 0.0973             & 108750.00         & $1.6883\times$       & $1.1982\times$    & $0.9889\times$    & $1.7376\times$    \\
        \texttt{specnorm-1000}     & 0.8710             & 147047.00         & $1.5239\times$       & $1.1336\times$    & $1.2688\times$    & $1.4862\times$    \\
        \texttt{list-4096}         & 0.0026             & 1537362.00        & $1562.9055\times$    & $1.6088\times$    & $1.6452\times$    & $3.4310\times$    \\
        \texttt{dbllist-4096}      & 0.0033             & 2061658.00        & $2742.9327\times$    & $1.5811\times$    & $1.9720\times$    & $3.1942\times$    \\
        \bottomrule
    \end{tabular}
    \label{tab:EvalGeneral}
\end{table}

\section{Limitations, Open Problems, and Future Work} \label{sec:Future}
The exact worst-case GCDS pause time--delay tradeoff is left open by our paper.
\pref{app:ReverseReduction} provides a reduction from GCDS to DRDS implying any
improved GCDS lower bounds would immediately give stronger DRDS bounds.
Hence, because lower bounds for the DRDS problem have resisted significant
improvement despite a large amount of research effort, it is unlikely that
significantly improved GCDS bounds can be found without novel lower bound
techniques.

Our analysis focused on collection delay, motivated by pathological cases where
modern GCs perform poorly when additional memory must be found quickly.
It would be interesting to explore other notions of garbage collection
performance and bounds on the performance of compacting collectors.
Many compacting collectors are susceptible to our bounds because they can be
converted into inplace collectors~\citep{baker_treadmill}, but we have not
formalized the extent of this connection.

We focused on asymptotic time complexity, ignoring memory overhead.
Our upper bound requires only a constant factor memory overhead, but the
constant is larger than for reference counting.
It would be interesting to determine a fine-grained tradeoff lower bound
between the constant factor memory overhead required and the time overhead.

\section{Related Work}
\label{sec:Related}
See~\citet{knuth_gc} for various garbage collection techniques and their
running time, and~\citet{omv} for a survey of recent bounds for dynamic graph
algorithms.

\paragraph{Early Garbage Collection Research}
The need for GC was noticed almost
immediately after the development of the linked list data
structure~\citep{ilp}. This led to a flurry of work on GC, including
reference counting~\citep{gelernter_gc,collins_gc},
mark-and-sweep~\citep{mccarthy_gc}, hybrids~\citep{mcbeth_cycles}, and copying
collectors~\citep{minsky_copying}.

Tradeoffs between pause times, predictability, and leakage in different GC
algorithms have been debated since these early days. \citet{weizenbaum_cycles}
publicized the failure of reference counting to avoid leaks in the presence of
cycles, but his solution had long pause times~\citep{mcbeth_cycles}.
\citet{weizenbaum_cycles2} suggested a hybrid collector, but pathological
programs causing poor collector performance remain to this day.
We provide a precise, formal proof showing some such tradeoffs are unavoidable.

\paragraph{Concurrent and Real-Time Garbage Collection}
Modern GC can run concurrently with the
program~\citep{baker_treadmill,copying_realtime,generational,minsky_copying,rtgc}.
Many are presented as so-called \emph{real-time} collectors, meaning under
certain well-defined scenarios they can guarantee pause times do not exceed a
certain constant limit and new allocations can always be
serviced~\citep{copying_realtime,baker_treadmill}.
Per the lower bounds in this paper, all such schemes are susceptible
to the sort of pathological examples described in~\pref{sec:Motivating}, where,
e.g., programs operating close to the memory limit must either reject
allocations unnecessarily or introduce long pauses.
For example, \pref{app:BakerCex} walks through an example program
for which a classic real-time collector would have to either reject allocations unnecessarily or
introduce large pause times.
In practice, this is avoided by overprovisioning memory, because many real-time
collectors can still make guarantees relating the peak logical memory usage to
the peak actual memory usage.
But overprovisioning resources is expensive, and difficult to do when memory
limits depend on other programs sharing resources.

\paragraph{Cyclic Reference Counting}
Many have attempted to modify reference counting to handle cycles without
requiring a full pass through the heap.  A key paper,~\citet{salkild}, seems to
be unavailable today so our discussion about it is based on second-hand reports
from other citations here.

\citet{brownbridge85} proposed one approach in 1985, but~\citet{salkild} soon
discovered it might free reachable memory. Apparently Salkild proposed a
solution that did not guarantee termination, and eventually \citet{ejh} had
found and proved correct a correct and terminating cyclic reference counting
scheme but no tight running time analysis was performed. Nowadays, algorithms based
on the local mark-scan of~\citep{localcrc} are more common in the literature,
such as~\citet{lazycrc} who removes the immediacy of cyclic reference
counting.  The similarity between such algorithms and traditional mark-scan is
observed by~\citet{unifiedtheory}.

As described in~\pref{sec:Sanity},
Sections~\ref{sec:Algorithm}~and~\ref{sec:LowerBounds2} answer a longstanding
open question in cyclic reference counting: can reference counting be modified
to always collect cycles, while guaranteeing reference counting-like overhead
when the heap is acyclic? \pref{sec:LowerBounds2} says \emph{no}: a logarithmic
slowdown is inevitable.  \pref{sec:Algorithm} says \emph{only} a logarithmic
slowdown is needed.

\paragraph{Languages Supporting Garbage Collection}
Popular implementations of the Java, Javascript, Lua, Python, and Go languages
all feature mature concurrent and/or generational mark-and-sweep
collectors~\citep{java_gc,js_gc,lua_gc,python_gc,go_gc}.
The Java language has been a particularly popular platform for GC
research~\citep{javagcsurvey1,javagcsurvey2,javagcsurvey3,javagcn}.
Garbage collectors also exist for the C and C++ languages~\citep{cgc1,cgc2}.

\paragraph{Research on New Collection Algorithms}
A particularly active area of modern research is in designing collectors that
work efficiently in the concurrent or distributed
setting~\citep{distgc1,distgc2,distgc3,distgc4} and in using different
heuristics than the standard generational approach~\citep{conngc}. Machine
learning might be able to improve GC performance in some common
cases~\citep{mlgc}.

\paragraph{Analyzing Garbage Collection}
Most authors seem to implicitly understand and take for granted that any collector will have pathological
cases with poor worst-case performance~\citep{knuth_gc,unifiedtheory}, and
focus instead on computing, e.g., the amount of memory needed to ensure the
limit is never reached~\citep{baker_treadmill,copying_realtime,rtgc}.
More common are qualitative or experimental comparisons between GC
schemes~\citep{unifiedtheory,javagcsurvey1}.

\paragraph{Evaluating and Handling Collector Tradeoffs}
Empirical issues with modern collectors are well known in the
literature~\citep{badgc1,badgc2}.
This paper complements that existing work with formal
limits on how much collectors may improve.
\citet{explaingc} have attempted to help debug GC performance issues.

\paragraph{Dynamic Reachability and Connectivity}
Dynamic connectivity (undirected graphs) and reachability (directed graphs) are
core problems in the study of dynamic graph algorithms. Many lower bounds are
known~\citep{lbdc,bestlbdc} in the cell probe model~\citep{cpm}. Nontrivial
upper bounds are known for \emph{undirected} dynamic connectivity, primarily
using Euler tour trees~\citep{xortrick,leveltrees}. The contribution of
our~\pref{sec:Algorithm} is to show that, beyond just undirected graphs,
\emph{acyclic} directed graphs can also be handled efficiently for our variant
of the problem.
\pref{sec:LowerBounds1} proves that the lower bounds for the general
reachability problem also apply to the more restricted GC problem.

\section{Conclusion}
\label{sec:Conclusion}
\emph{Collection delay} is the worst-case time between a memory region first
becoming unreachable in the heap and the GC collecting it.
Modern real-time GC algorithms accept suboptimal collection delay in some
pathological cases in order to guarantee constant-sized pause times when memory
is abundant.
This motivates asking whether modern GCs can be modified to reduce the
worst-case delay without significantly increasing pause times.
We provide a formal proof that extreme improvement in the worst case is not
possible: superlogarithmic collection delay is inevitable for some pathological
programs unless longer pause times are allowed.
Our results hold for any GC implementing a formal mutator-observer style
interface defined in this paper.
Our proofs work via a nontrivial connection to fundamental data structures
lower bounds, hence any stronger GC lower bound than ours would lead
immediately to an improvement in important data structure lower bounds and
vice-versa.
We also describe a GC with some interesting asymptotic behavior, although it
introduces too much overhead to recommend in non-pathological settings.
\clearpage

\section*{Acknowledgements}
I would like to thank the anonymous reviewers, whose suggestions have
dramatically improved the quality of the paper;
Geoff Ramseyer, Alex Ozdemir, David K.\ Zhang, and Scott Kovach for extended
discussions improving this work;
as well as Dawson Engler, Zachary Yedidia, Akshay Srivatsan, members of the
Stanford theory group, and attendees of the Stanford software lunch, who
provided helpful insights, conversations, and proofreading.


\section*{Data Availability Statement}
Implementations of different GCs in the Lua interpreter (as evaluated
in~\pref{sec:Algorithm}) are available at
\url{https://doi.org/10.5281/zenodo.14942311}~\citep{artifact}.

\section*{Full Version With Appendices}
The full version of this paper, with appendices, is available at
\url{https://doi.org/10.5281/zenodo.14948284}.

\section*{Funding Statement}
This work was generously funded via grants NSF DGE-1656518 and Stanford IOG
Research Hub 281101-1-UDCPQ 298911.

\bibliographystyle{ACM-Reference-Format}
\bibliography{main}

\appendix
\clearpage
\section{Extended Motivating Examples}
\label{app:Motivating}
Full code for the motivating examples from~\pref{sec:Motivating} is provided
below.
We also describe modified versions that would cause similarly pathological
behavior on GCs supporting generational and/or reference counting extensions to
their collectors.

\subsection{GC Thrashing}
The full Lua code is given below.
We simulate large allocation blobs via descriptors, and keep track of the
number of allocated blobs that have not yet been collected.
This tracks logical memory, ignoring the overhead of the interpreter itself.
This is likely what a user not interested in modifying the interpreter would
end up implementing.
It would also be reasonable to implement the blobs as allocations on a
different machine, or reservations on some external resource (e.g., open
network connections).

\begin{minted}{Lua}
START_TIME = os.clock()

TOTAL_ALLOC = 0
N_ITEMS = 1000000

ALLOC_SIZE = 1
ALLOC_BUFFER = 1000

ALLOC_LIMIT = (ALLOC_SIZE * N_ITEMS) + ALLOC_BUFFER
-- arg[1] can be "limit", "nolimit", or "manual"
ENFORCE_LIMIT = (arg[1] ~= "nolimit")
MANUAL_FREE = (arg[1] == "manual")

local function clock(message)
    local time = os.clock()
    local lpad = 20
    if #message < lpad then
        message = string.rep(" ", lpad - #message) .. message
    end
    print(message, "@", time - START_TIME, "m", TOTAL_ALLOC)
    -- update the START_TIME to ignore time taken printing the clock message
    START_TIME = START_TIME + (os.clock() - time)
end

local function collect_blob(self)
    TOTAL_ALLOC = TOTAL_ALLOC - self.fake_alloc_size
end

local function fresh_blob()
    -- simulate an allocation
    if ENFORCE_LIMIT and TOTAL_ALLOC >= ALLOC_LIMIT then
        clock("Start collection")
        collectgarbage("collect")
        clock("End collection")
        if TOTAL_ALLOC >= ALLOC_LIMIT then
            print("OUT OF MEMORY!")
            os.exit(1)
        end
    end
    TOTAL_ALLOC = TOTAL_ALLOC + ALLOC_SIZE
    local res = setmetatable({fake_alloc_size=ALLOC_SIZE}, {__gc=collect_blob})
    return res
end

local function free_blob(blob)
    TOTAL_ALLOC = TOTAL_ALLOC - blob.fake_alloc_size
    blob.fake_alloc_size = 0
    setmetatable(blob, {})
end

local function fetch_item()
    return {data=fresh_blob()}
end

local function process_item(item)
    local work_space = fresh_blob()
    if MANUAL_FREE then
        free_blob(work_space)
    end
    return 1
end

-- stage 1: fetch a lot of items to process
local items = {}
for i=1,N_ITEMS do
    items[i] = fetch_item()
    clock("Iter")
end
clock("STAGE")
-- stage 2: process each item
local summary = 0
for i=1,N_ITEMS do
    summary = summary + process_item(items[i])
    clock("Iter")
end
clock("STAGE")
-- Don't bother GC'ing at the end
io.flush()
os.exit(0)
\end{minted}

\subsection{GC Thrashing: Breaking Reference Counting and Generations}
The above code performs well under both generational and reference
counting-based collectors.
However, it can be easily modified to make reference counting useless on the
example: simply make \texttt{fresh\_blob()} create and return a node from a
cyclicly linked list.
To cause similar behavior on a language using generational collection, the
second stage could swap the blob allocated in \texttt{process\_item} with
one of the blobs allocated in the first stage.
Then it will be the older allocation region that becomes
unreachable, hence generational collection will not help.

\subsection{Delayed Finalization}
The delayed finalization server and client scripts are given below.
\begin{minted}{Lua}
------------- SERVER
N_SLOTS=100

local slot_open = {}
for i=1,N_SLOTS do slot_open[i] = true end

local worklist = {}

os.execute("mkdir -p work")
os.execute("rm -f work/*")

local function path_exists(path)
  local f, _, _ = io.open(path, "r")
  if f then
    f:close()
    return true
  end
  return false
end

local function finalize(descriptor)
  os.remove(descriptor.path)
  slot_open[descriptor.i] = true
end

local function poll()
  -- phase 1: poll for more work
  for i=1,N_SLOTS do
    if slot_open[i] then
      local fname = "work/" .. tostring(i) .. ".txt"
      if path_exists(fname) then
        slot_open[i] = false
        table.insert(worklist,setmetatable({i=i,path=fname},{__gc=finalize}))
      end
    end
  end
end

local function process(item)
    print("Processing ... " .. tostring(item.i))
    item = nil
end

local function drain()
  -- phase 2: drain the work queue
  while #worklist > 0 do
    process(table.remove(worklist))
  end
end

while true do
  poll()
  drain()
end

------------- CLIENT (different file)
-- poll, waiting for our slot to open up. when it does, fill it.

TARGET=10000
MY_SLOT=1

local count = 0
while count < TARGET do
  local fname = "work/" .. tostring(MY_SLOT) .. ".txt"
  local f, _, _ = io.open(fname, "r")
  if not f then
    local f = assert(io.open("work/tmp.txt", "w"))
    f:write("Hello, World!\n")
    f:flush()
    f:close()
    os.rename("work/tmp.txt", fname)
    count = count + 1
    print(count)
  else
    f:close()
  end
end
\end{minted}

\subsection{Delayed Finalization: Breaking Reference Counting and Generations}
Here again, reference counting would not help if each workitem were made part
of a cyclic list.
Generational collection would also fail to help if, e.g., it takes enough time
to process workitems that they become part of an older generation.

\clearpage
\section{Counterexample to Claim of~Pepels~et.~al.}
\label{app:PepelsCex}
\citet{ejh} claims that their cyclic reference counting scheme ``has
no time overhead compared to classical reference counting algorithms, if there
are no cycles in the graph.'' This claim would directly contradict our lower
bound~(\pref{thm:AcyclicLB}). Unfortunately, no extended justification for the
claim is given in the text. We believe that the claim is a mistake in the
English translation.

We now present a sequence of operations that causes their algorithm to take
asymptotically more time than reference counting, even though there are no
cycles in the graph, hence directly refuting their claim.

First, construct the following heap, noticing that all pointers must be
assigned strong because the heap is treeshaped:
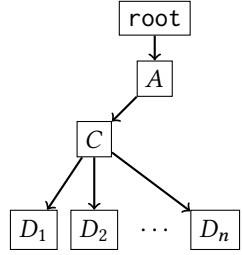
\begin{center}
    \begin{tikzpicture}[scale=0.8]
            \node[gcnode] (Root) at (0, 1) {$\GCRoot$};
            \node[gcnode] (A) at (0, 0) {$A$};
            \node[gcnode] (C) at (-1, -1) {$C$};
            \node[gcnode] (D1) at (-2, -2.5) {$D_1$};
            \node[gcnode] (D2) at (-1, -2.5) {$D_2$};
            \node (DDots) at (0, -2.5) {$\ldots$};
            \node[gcnode] (Dn) at (1, -2.5) {$D_n$};

            \draw[thick] (Root) edge[->] (A);
            \draw[thick] (A) edge[->] (C);
            \draw[thick] (C) edge[->] (D1);
            \draw[thick] (C) edge[->] (D2);
            \draw[thick] (C) edge[->] (Dn);
    \end{tikzpicture}
\end{center}

Now, allocate a new memory region $B$ and copy the $A \to C$ pointer to $B \to
C$. Notice that the \texttt{CopyPtr} method in~\citet{ejh} makes it weak, so we
have the following heap:
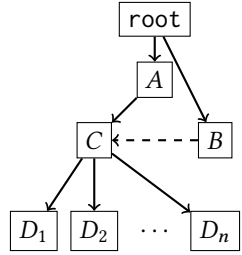
\begin{center}
    \begin{tikzpicture}[scale=0.8]
            \node[gcnode] (Root) at (0, 1) {$\GCRoot$};
            \node[gcnode] (A) at (0, 0) {$A$};
            \node[gcnode] (B) at (1, -1) {$B$};
            \node[gcnode] (C) at (-1, -1) {$C$};
            \node[gcnode] (D1) at (-2, -2.5) {$D_1$};
            \node[gcnode] (D2) at (-1, -2.5) {$D_2$};
            \node (DDots) at (0, -2.5) {$\ldots$};
            \node[gcnode] (Dn) at (1, -2.5) {$D_n$};

            \draw[thick] (Root) edge[->] (A);
            \draw[thick] (Root) edge[->] (B);
            \draw[thick] (A) edge[->] (C);
            \draw[thick] (B) edge[->,dashed] (C);
            \draw[thick] (C) edge[->] (D1);
            \draw[thick] (C) edge[->] (D2);
            \draw[thick] (C) edge[->] (Dn);
    \end{tikzpicture}
\end{center}

Finally, call \texttt{DeletePtr} on the edge $A \to C$. Following the
definition of the \texttt{DeletePtr} method in~\citet{ejh}, it, among other
things:
\begin{enumerate}
    \item Calls \texttt{AdjustPresumableCycle} on $C$, which:
        \begin{enumerate}
            \item Calls \texttt{Swap} on $C$, making $B \to C$ strong,
            \item Calls \texttt{AdjustGraphBetween} on $C, C$, which:
                \begin{enumerate}
                    \item Iterates over the strong-pointer sons of $C$.
                \end{enumerate}
        \end{enumerate}
\end{enumerate}
But at that point, $C$ has $n$ strong-pointer sons, namely $D_1$ through $D_n$.
Hence the \texttt{DeletePtr} operation will take $O(n)$ time, even though the
heap has always been acyclic and reference counting would have only taken
$O(1)$ time.

\clearpage
\section{Layered Permutation Graphs}
\label{app:LPRDS}
\pref{sec:LPGCDS} formalized the LPRDS slightly differently from the original
in~\citet{lbdc}.
They phrase their lower bound in terms of (using our terminology) the general
undirected dynamic connectivity data structure problem (DCDS, identical
to~\pref{def:DRDS} but for undirected graphs).
However, the sequence of operations that they prove requires $\Omega(\log{n})$
time satisfies the conditions of the LPRDS we defined~(\pref{def:LPRDS}).
In particular, their sequence of operations always involves permuting one layer
of edges at a time, then querying for connectedness from each node in the input
layer to some node in a later layer.

To show that lower bounds from~\citet{lbdc} apply to the LPRDS problem we
defined in~\pref{def:LPRDS}, it suffices to show that an efficient LPRDS could
be used to construct an undirected DCDS that has identical amortized time
complexity to the LPRDS when the queries are in the form used by~\citet{lbdc},
i.e., supported by the LPRDS.
Such a reduction is shown in~\pref{fig:LPAlgs}.

The basic idea is to wrap the LPRDS in another data structure. As we get graph
updates, we store them in a na\"ive graph representation.
When the first connectedness query is made, we perform a DFS to determine if
the graph is of the LPRDS form;\footnote{The results of~\citet{lbdc} apply in an
amortized setting so we can amortize away the cost of the DFS.} if so, we
assign each node its proper layer number, initialize the LPRDS, and then
proceed by querying the LPRDS.
If the LPRDS invariants are ever violated, we go back to the na\"ive graph
algorithm.

\newcommand\inited{\mathrm{use\_lprds}}
\begin{figure}[t]
    \begin{minipage}{0.54\textwidth}
        \begin{algorithm}[H]
            \caption{$\UDCInsert(v_i \uto v_j)$} \label{alg:lpinsert}
            Rename $i, j$ so $L(v_i) \leq L(v_j)$\;
            $E(v_i \uto v_j) \gets E(v_i \uto v_j) + 1$\;
            $\inited{} \gets \inited{} \wedge L(v_i) = L(v_j) - 1$\;
            $\inited{} \gets \inited{} \wedge \mathrm{indeg}(v_j) \leq 1$\;
            $\inited{} \gets \inited{} \wedge \mathrm{outdeg}(v_i) \leq 1$\;
            \If{$\inited{}$}{
                $\LPCInsert(v_i \to v_j)$\;
            }
        \end{algorithm}
    \end{minipage}
    \hfill
    \begin{minipage}{0.45\textwidth}
        \begin{algorithm}[H]
            \caption{$\UDCInit()$} \label{alg:lpinit}
            $E, L, \inited{}, R \gets \emptyset, \emptyset, ?, \emptyset$\;
        \end{algorithm}
        \begin{algorithm}[H]
            \caption{$\UDCDelete(v_i \uto v_j)$} \label{alg:lpdelete}
            Rename $i, j$ so $L(v_i) \leq L(v_j)$\;
            $E(v_i \uto v_j) \gets E(v_i \uto v_j) + 1$\;
            \If{$\inited{}$}{
                $\LPCDelete(v_i \to v_j)$\;
            }
        \end{algorithm}
    \end{minipage}
    \begin{algorithm}[H]
        \caption{$\UDCConnected(v_i \uto^+ v_j)$} \label{alg:lpquery}
        $R \gets R \cup \{ v_i \}$\;
        \If{$\inited{} = ? \wedge \abs{R} = n$}{
            \eIf{DFS from $R$ finds graph is $n$ disjoint paths}{
                \ForAll{vertex $v$}{
                    $L(v) \gets$ distance of $v$ from a node in $R$\;
                }
                Initialize the LPRDS with layers from $L$ and edges from $E$\;
                $\inited{} \gets \top$\;
            }{
                $\inited{} \gets \bot$\;
            }
        }
        \lIf{$\inited{} = \top \wedge v_i \in R$}{ \Return{$\LPCConnected(v_i \to^+ v_j)$} }
        \Return{BFS or DFS connectedness search}\;
    \end{algorithm}
    \caption{Showing our directed LPRDS can be used to speed up a general DRDS
    when the graph is of the form used by~\citet{lbdc}.}
    \label{fig:LPAlgs}
\end{figure}

\clearpage
\section{Pathological Example for Baker's Treadmill}
\label{app:BakerCex}
Baker's treadmill algorithm~\citep{baker_treadmill,copying_realtime} is a
paradigmatic example of a real-time garbage collector. All regions except
$\GCRoot$ are initially marked white; $\GCRoot$ is marked gray. During each
collector operation, some constant number of gray node(s) are turned black
and their white children are made gray. When all nodes are either black or white,
i.e., none are gray, we know the white nodes are unreachable and we may free
them. Finally, the black nodes are marked as white, $\GCRoot$ is marked
gray, and the process repeats.

Baker~\citep{baker_treadmill,copying_realtime} proves that \emph{if the program satisfies a number of conditions}
regarding, e.g., how frequently memory is allocated vs.\ released, then the
collector will always make enough progress to ensure at least one free memory
region is known by the allocator whenever the program requests memory.
Unfortunately, there exist programs that force Baker's treadmill to either
introduce superconstant pause times, i.e., breaking the real-time guarantee, or
incorrectly report that there is no available memory.
This section briefly describes one such program.

The program first fills up the available memory with a single large linked
list, preserving a pointer to the second-to-last node in the list:
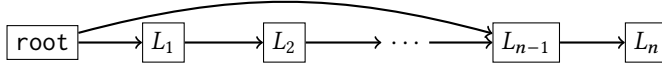
\begin{center}
    \begin{tikzpicture}[scale=0.8]
            \node[gcnode] (Root) at (0, 0) {$\GCRoot$};
            \node[gcnode] (L1) at (2, 0) {$L_1$};
            \node[gcnode] (L2) at (4, 0) {$L_2$};
            \node (Dots) at (6, 0) {$\ldots$};
            \node[gcnode] (LN1) at (8, 0) {$L_{n-1}$};
            \node[gcnode] (LN) at (10, 0) {$L_n$};

            \draw[thick] (Root) edge[->] (L1);
            \draw[thick] (Root) edge[->,bend left=15] (LN1);
            \draw[thick] (L1) edge[->] (L2);
            \draw[thick] (L2) edge[->] (Dots);
            \draw[thick] (Dots) edge[->] (LN1);
            \draw[thick] (LN1) edge[->] (LN);
    \end{tikzpicture}
\end{center}

Assume that after building this heap the collector completes a flip, i.e., one
complete pass of the collector through the heap. This can be triggered either
by requesting a new memory region or by performing no-ops until the
mark-and-sweep process is completed.

In any case, the heap now has $\GCRoot$ marked in gray and all other nodes in
white. Suppose the program now deletes the $L_{n-1} \to L_n$ pointer. This
makes $L_n$ unreachable, but the collector is unable to guarantee that until it
finishes marking the \emph{entire} rest of the heap $L_1, \ldots, L_{n-1}$.
Hence, if a new allocation is requested immediately after deleting the edge to
$L_n$, the collector will be forced to either complete the $O(n)$-time
mark-and-sweep pass or report that it cannot find a free region, even though
$L_n$ is indeed unreachable. This scenario shows how pathological programs can
cause non-real-time behavior if they operate too close to the memory limit.

Note in this particular scenario, combining the treadmill with a reference
counting collector would enable quick collection. However, our lower bounds
guarantee pathological examples will always exist: e.g., $L_n$ could be the
head of a cycle in the graph, in which case na\"ive reference counting would not
help.

\section{Lua Experiment Details}
\label{app:ExperimentDetails}
We forked the standard Lua 5.4.6 interpreter, removing its garbage collection
code and inserting hooks where Lua VM instructions manipulate Lua references.
These hooks call $\GCAllocate$, $\GCInsert$, and $\GCDelete$ methods of the
underlying GCDS.
We wrote two GCDS implementations, one for our GCDS and the other for SynCC
(prior work~\citep{bacon_crc}).

\paragraph{Implementation of our GCDS}
Our GCDS implementation is based on the algorithm in~\pref{sec:Algorithm}.
We used splay trees to implement the Euler tour trees.

\paragraph{Implementation of SynCC}
We compared to the SynCC approach of~\citet{bacon_crc}.
Later in their paper~\citet{bacon_crc} describe how it can be used in a delayed
fashion but we implemented it as it is first presented, i.e., a collector with
$d(n) = 1$, to compare against our GCDS.
$\GCAllocate$, $\GCInsert$, and $\GCDelete$ maintain a copy of the points-to
graph.
After deletion of an edge, we apply the algorithm in~\citet{bacon_crc} to
determine whether the region is reachable or, if not, to identify all regions
that can be freed.

\paragraph{Memory Overhead}
For both implementations we made an effort to reduce memory overhead where
obvious options existed, e.g., using bitfields for boolean flags, but focused
on implementation simplicity and time overhead first and foremost.
In particular, for implementation simplicity we maintained a copy of the
points-to graph in a separate graph data structure.
The default Lua collector embeds this graph within the elements themselves,
which allows it to achieve better memory overhead.
Unfortunately, that approach requires type-specific iterators, hence
complicating implementation of the GC, hence our simpler but
higher-memory-usage approach.

\paragraph{General Benchmarks Used}
All of the general benchmarks~(\pref{sec:EvalGeneral}) except for
\texttt{list-4096} and \texttt{dbllist-4096} are from the Programming Languages
Benchmarks project~\citep{plb}. We excluded the \texttt{coro-prime-sieve}
benchmark, as it caused a stack overflow even on the unmodified Lua interpreter
when run on inputs large enough to produce nontrivial time.
The \texttt{list-4096} benchmark creates a large linked list, iterates over it
to compute its length, and then deletes it.
The \texttt{dbllist-4096} benchmark is identical, except it builds a doubly
linked list.

\section{Reduction from GCDS to DRDS}
\label{app:ReverseReduction}
Most of the paper proved that an efficient GCDS could be used to build an
efficient DRDS; this section proves the converse.
The result of this proof is to show that discovering improved bounds on the
GCDS problem is just as hard as discovering improved bounds on the DRDS
problem.

\begin{figure}[t]
    \begin{minipage}{0.49\textwidth}
        \begin{algorithm}[H]
            \caption{$\GCInsert(v_i \to v_j)$} \label{alg:lbcinsert}
            $\DCInsert(v_i \to v_j)$\;
            $E(v_i \to v_j) \gets E(v_i \to v_j) + 1$\;
        \end{algorithm}
        \begin{algorithm}[H]
            \caption{$\GCAllocate()$} \label{alg:lbcdelete}
            $v_i \gets$ fresh node\;
            $\DCInsert(\GCRoot \to v_i)$\;
            \Return{$v_i$}\;
        \end{algorithm}
    \end{minipage}
    \hfill
    \begin{minipage}{0.49\textwidth}
        \begin{algorithm}[H]
            \caption{$\GCDelete(v_i \to v_j)$} \label{alg:lbcquery}
            $\DCDelete(v_i \to v_j)$\;
            \If{$\neg\DCConnected(\GCRoot \to^+ v_j)$}{
                $D \gets \{ v_j \}$\;
                \ForAll{$E(v_j \to m) > 0$}{
                    $D \gets D \cup \GCDelete(v_j \to m)$\;
                }
                FreeList $\gets$ FreeList $\cup D$\;
            }
        \end{algorithm}
    \end{minipage}
    \begin{algorithm}[H]
        \caption{$\GCStep()$} \label{alg:lbcfindone}
        \textbf{no-op}\;
    \end{algorithm}
    \caption{Reduction from GCDS to DRDS.}
    \label{fig:LBCAlgs}
\end{figure}

\begin{theorem}
    \label{thm:ReverseReduction}
    Assume the existence of an $O(g(n))$-time DRDS.
    Then there exists a GCDS with $d(n) = 1$ and $\tAR(n) = O(g(n))$.
\end{theorem}
\begin{proof}
    The reduction is shown in~\pref{fig:LBCAlgs}.
    When a pointer is deleted we use the DRDS as an oracle to check whether
    there remains any path from $\GCRoot$ to the node; if not, we delete it and
    recurse on its outgoing pointers similar to reference counting.
\end{proof}

\section{GC to GCDS}
\label{app:GC2GCDS}
Having a formal model is unavoidable when proving rigorous bounds; to prove
bounds on `garbage collection' we must clearly delineate what behaviors we
require of `garbage collection.'
We designed the GCDS interface in~\pref{def:GCDS} to match the standard
mutator-observer interface provided by existing inplace GCs for imperative
programs.
It is interesting to ask whether there might be an imperative programming
language using a GC that can \emph{not} be made to fit this interface without
performance degradation.
With some caveats (below), the answer is no.

Given access to a general-purpose, imperative programming language like Python
or Lua with a hypothetical efficient GC, you could simulate a GCDS by making
$\GCAllocate$ create a new set object for the memory region and simulating
$\GCInsert(a\to b)$ and $\GCDelete(a \to b)$ by adding and
removing references to $b$ from the set in memory region $a$.
Memory regions can be annotated with finalizers that add them to $\GCFreeList$.

\paragraph{Caveat 1: Need for Finalizers}
The sketch above requires the language support finalizers so that the
simulated GCDS can determine when a region is collectable, i.e., can be
added to $\GCFreeList$.
Some sort of feedback from the GC to the application seems necessary, as
otherwise the GC could simply never free anything.
Perhaps the simplest GC interface would require the ability to set a memory
limit and then query whether a new allocation can be made or not.
A language supporting such a GC could be used to implement a GCDS variant where
instead of a free list the GCDS allows querying only whether or not anything
has been collected.
This GCDS variant would be sufficient to carry out our reductions
in~\pref{sec:LowerBounds1} and~\pref{sec:LowerBounds2}, so our results still
hold in this setting.

\paragraph{Caveat 2: Nonlocal Modifications}
The final caveat is that the GCDS allows operations involving any nodes
reachable from $\GCRoot$, but in an imperative program, each operation can
usually only write to a region reachable within a constant number of pointer
indirections from a local or global variable.
This corresponds to adding the additional restriction in~\pref{def:GCDS} that
all operations must involve nodes at most some bounded distance from $\GCRoot$.
Thankfully, our reduction in~\pref{sec:LowerBounds1} satisfies this
requirement, and our reduction in~\pref{sec:LowerBounds2} can be made to
satisfy this requirement by introducing an auxiliary node similar to the $X$
node used in~\pref{sec:LowerBounds1}.
Hence, our results still hold in this setting as well.

\section{Rigorous Definitions for GC Performance}
\label{app:FormalDefinitions}
In this section we give more rigorous definitions for the notions of GCDS delay
and running time sketched in~\pref{sec:DefiningTime}.
We will assume here a deterministic GCDS; see~\pref{sec:AltDefinitions} for
discussion of nondeterministic GCDS.
The first key idea is the \emph{associated heap multigraph} of a sequence of
GCDS operations, which associates with each sequence of GCDS operations the
state of the heap that those operations describes.
Notably, the GCDS is not required to actually store the associated heap
multigraph (although it is frequently helpful, see~\pref{sec:GCDSExamples};
rather, the associated heap multigraph is a theoretical construction that we
use in our definitions of GCDS performance.

\begin{definition}
    Let $p_1, \ldots, p_n$ be a sequence of GCDS operations~(\pref{def:GCDS})
    and $r_1, \ldots, r_n$ be the values returned by the GCDS for each of those
    operations in the sequence (or $\bot$ if the operation returns nothing).
    Then, the \emph{associated heap multigraph} $\eta(p_1, \ldots, p_n)$ is a
    directed multigraph (i.e., directed graph with possibly multiple edges
    between the same pair of nodes) defined inductively:
    \begin{enumerate}
        \item If $n = 0$, then $\eta(\emptyset)$ is a graph with a single node,
            named $\GCRoot$, and no edges.

        \item If $p_n$ is $\GCAllocate()$, then $\eta(p_1, \ldots, p_n)$ is
            $\eta(p_1, \ldots, p_{n-1})$ with a new node $r_n$ added to it,
            along with a new edge $\GCRoot \to r_n$.

        \item If $p_n$ is $\GCInsert(a \to b)$, then $\eta(p_1, \ldots, p_n)$
            is $\eta(p_1, \ldots, p_{n-1})$ with a new edge $a \to b$ (if $a$
            and $b$ are nodes in the graph).

        \item If $p_n$ is $\GCDelete(a \to b)$, then $\eta(p_1, \ldots, p_n)$
            is $\eta(p_1, \ldots, p_{n-1})$ with one edge $a \to b$ removed (if
            any exist).

        \item If $p_n$ is $\GCStep()$, then $\eta(p_1, \ldots, p_n)$
            is $\eta(p_1, \ldots, p_{n-1})$.
    \end{enumerate}
\end{definition}

Recall from~\pref{def:GCDS} that a GCDS could make some reasonable assumptions
about its inputs, e.g., that no node is used unless it is reachable from
$\GCRoot$.
We now define precisely the meaning of a sequence being \emph{valid}.
In particular, GCDS behavior need not be defined on invalid sequences.
\begin{definition}
    Let $p_1, \ldots, p_n$ be a sequence of GCDS operations~(\pref{def:GCDS}).
    The sequence is \emph{valid} if it satisfies the following inductive
    definition.
    \begin{enumerate}
        \item If $n = 0$, then the sequence is valid.

        \item If $p_n$ is $\GCAllocate()$ or $\GCStep()$, then $p_1, \ldots,
            p_n$ is valid if $p_1, \ldots, p_{n-1}$ is valid.

        \item If $p_n$ is $\GCInsert(a \to b)$, then $p_1, \ldots, p_n$ is
            valid if (1) $\eta(p_1, \ldots, p_{n-1})$ is valid, and (2)
            $\GCRoot$ can reach $a$ and $b$ in $\eta(p_1, \ldots, p_{n-1})$.

        \item If $p_n$ is $\GCDelete(a \to b)$, then $p_1, \ldots, p_n$ is
            valid if (1) $\eta(p_1, \ldots, p_{n-1})$ is valid, (2)
            $\GCRoot$ can reach $a$ and $b$ in $\eta(p_1, \ldots, p_{n-1})$,
            and (3) there is at least one edge $a \to b$ in $\eta(p_1, \ldots,
            p_{n-1})$.
    \end{enumerate}
\end{definition}

We can now define more rigorously the \emph{worst-case delay} $d(n)$, which
in~\pref{def:Delay} we described as ``the pointwise minimal function such that,
in any sequence of GCDS operations involving at most $n$ nodes (i.e., making at
most $n$ calls to $\GCAllocate$), if some operation makes a node $u$
unreachable, $u$ is added to $\GCFreeList$ after at most $d(n)$ more GCDS
operations (including the operation that makes it unreachable).''
\begin{definition}
    Let $S_n$ be the set of valid nonempty GCDS sequences $p_1, \ldots, p_k$
    such that at most $n$ of the $p_i$s are $\GCAllocate()$s.
    Consider some particular GCDS, and let $F(p_1, \ldots, p_i)$ be the set of
    all nodes added by the GCDS to $\GCFreeList$ during the sequence of
    operations $p_1, \ldots, p_i$.
    Then the \emph{worst-case delay} $d(n)$ of the GCDS is defined to be:
    \[
        \begin{aligned}
            d(n) \coloneqq
            \mathrm{max}(\{ \delta \quad \mid \quad
            &\text{exist $p_1, \ldots, p_k \in S_n$,
            $t \leq k - \delta + 1$, and
            $a \in \eta(p_1, \ldots, p_{t-1})$ such that} \\
            &\text{(1) there is a path $\GCRoot \to^+ a$ in $\eta(p_1, \ldots, p_{t-1})$,} \\
            &\text{(2) there is not a path $\GCRoot \to^+ a$ in $\eta(p_1, \ldots, p_t)$, and} \\
            &\text{(3) $a \not\in F(p_1, \ldots, p_{t-1+\delta})$} \}).
        \end{aligned}
    \]
    If there is no maximum, we write $d(n) = \infty$.
\end{definition}
Intuitively, this definition is looking at sequences $p_1, \ldots, p_k$ and
positions $t$ in that sequence such that $a$ became unreachable on operation
$p_t$, and $a$ has not been added to the free list by operation
$p_{t-1+\delta}$.
The largest such $\delta$, over all sequences involving at most $n$ nodes, is
called $d(n)$.

We can also define the \emph{worst-case pause time} of a GCDS, which we earlier
described as ``the pointwise minimal function such that, in any sequence
of GCDS operations involving at most $n$ nodes, each GCDS operation in the
sequence takes time at most $t(n)$.''
\begin{definition}
    \label{def:PrecisePause}
    Let $S_n$ be the set of valid nonempty GCDS sequences $p_1, \ldots, p_k$
    such that at most $n$ of the $p_i$s are $\GCAllocate()$s.
    Consider some particular GCDS and let $T(p_1, \ldots, p_k)$ be the time
    taken by the GCDS to process the valid sequence of operations $p_1, \ldots,
    p_k$; in particular, $T(p_1, \ldots, p_k) - T(p_1, \ldots, p_{k-1})$ is the
    marginal time to process $p_k$.
    Then the \emph{worst-case pause time} is
    \[
        \begin{aligned}
            t(n) \coloneqq \mathrm{max}(\{ T(p_1, \ldots, p_k) - T(p_1, \ldots, p_{k-1}) \mid
            p_1, \ldots, p_k \in S_n \}).
        \end{aligned}
    \]
\end{definition}
Note in particular that $S_n$ is closed under taking (nonempty) prefixes, so
the above definition also captures the time taken for every subsequence of
operations.

We can now define $\tAR$, $\tAAR$, and $\tSAAR$.
Each one is very similar to the definition of worst-case pause time, except
taking the maximum over an increasing small subset of sequences.
\begin{definition}
    Using the definitions from~\pref{def:PrecisePause}, the
    \emph{all-reachable} worst-case pause time $\tAR(n)$ of a GCDS is
    restricted to sequences that never make any node unreachable:
    \[
        \begin{aligned}
            \tAR(n) \coloneqq
            \mathrm{max}(\{ &T(p_1, \ldots, p_k) - T(p_1, \ldots, p_{k-1}) \mid
            \text{exists $p_1, \ldots, p_k \in S_n$ such that, for every $i \leq k$,} \\
            &\text{(1) every node in $\eta(p_1, \ldots, p_i)$ is reachable from $\GCRoot$} \}).
        \end{aligned}
    \]
\end{definition}
\begin{definition}
    Using the definitions from~\pref{def:PrecisePause}, the \emph{acyclic all-reachable} worst-case pause time $\tAAR(n)$ of a
    GCDS is restricted to sequences that never make any node unreachable and
    never introduce a cycle:
    \[
        \begin{aligned}
            \tAAR(n) \coloneqq
            \mathrm{max}(\{ &T(p_1, \ldots, p_k) - T(p_1, \ldots, p_{k-1}) \mid
            \text{exists $p_1, \ldots, p_k \in S_n$ such that, for every $i \leq k$,} \\
            &\text{(1) every node in $\eta(p_1, \ldots, p_i)$ is reachable from $\GCRoot$, and} \\
            &\text{(2) the graph $\eta(p_1, \ldots, p_i)$ is acyclic}
            \}).
        \end{aligned}
    \]
\end{definition}

The definition of sparse acyclic all-reachable worst-case pause time is
fundamentally asymptotic due to the definition of sparsity; we first define a
corresponding nonasymptotic notion of \emph{$m$-sparse} acyclic all-reachable
worst-case pause time.
\begin{definition}
    Using the definitions from~\pref{def:PrecisePause}, the \emph{$m$-sparse
    acyclic all-reachable} worst-case pause time $\tkSAAR(n)$ is restricted to
    sequences that never make any node unreachable, never introduce a cycle,
    and never have more than $m$ edges leaving any node.
    \[
        \begin{aligned}
            \tkSAAR(n) \coloneqq
            \mathrm{max}(\{ &T(p_1, \ldots, p_k) - T(p_1, \ldots, p_{k-1}) \mid
            \text{exists $p_1, \ldots, p_k \in S_n$ such that, for every $i \leq k$,} \\
            &\text{(1) every node in $\eta(p_1, \ldots, p_i)$ is reachable from $\GCRoot$, and} \\
            &\text{(2) the graph $\eta(p_1, \ldots, p_i)$ is acyclic, and} \\
            &\text{(3) all nodes in $\eta(p_1, \ldots, p_i)$ have outdegree at most $m$}
            \}).
        \end{aligned}
    \]
\end{definition}
Finally, $\tSAAR$ is defined as the asymptotic growth rate of the $\tkSAAR$s,
if an upper bound on the growth rate exists independent of $m$ (otherwise we
write $\tSAAR(n) = \infty$.
\begin{definition}
    A GCDS has \emph{sparse acyclic all-reachable} worst-case pause time
    $\tSAAR(n)$ if for every constant $m$, $\tkSAAR(n) = O(\tSAAR(n))$.
\end{definition}

\section{Variants of Definitions Our Bounds Work For}
\label{sec:AltDefinitions}
We now briefly discuss two alternate definitions that our bounds also work for,
and how using these alternate definitions makes the bounds apply to other GCDS
implementations.

\subsection{First-Delay}
Above, we defined delay to be the maximum number of operations between when a
node becomes unreachable and when it is added to the free list.
But our reductions actually only rely on the ability to detect when
\emph{anything} has been made unreachable, i.e., we only really need to look at
the difference between \emph{the first node becoming unreachable} in a sequence
and when any node gets added to the free list.
This leads to the following definition of \emph{first delay}.
\begin{definition}
    Let $S_n$ be the set of valid nonempty GCDS sequences $p_1, \ldots, p_k$
    such that at most $n$ of the $p_i$s are $\GCAllocate()$s.
    Consider some particular GCDS, and let $F(p_1, \ldots, p_i)$ be the set of
    all nodes added by the GCDS to $\GCFreeList$ during the sequence of
    operations $p_1, \ldots, p_i$.
    Then the \emph{worst-case first delay} $d_1(n)$ of the GCDS is defined to be:
    \[
        \begin{aligned}
            d_1(n) \coloneqq
            \mathrm{max}(\{ \delta \quad \mid \quad
            &\text{exists $p_1, \ldots, p_k \in S_n$ and
            $t \leq k - \delta$ such that} \\
            &\text{(1) $\GCRoot$ can reach every node in $\eta(p_1, \ldots, p_{t-1})$,} \\
            &\text{(2) there is some node that $\GCRoot$ cannot reach in $\eta(p_1, \ldots, p_t)$, and} \\
            &\text{(3) $F(p_1, \ldots, p_{t-1+\delta})$ is empty} \}).
        \end{aligned}
    \]
\end{definition}
The results in~\pref{sec:LowerBounds1} and~\pref{sec:LowerBounds2} all still
hold when worst-case delay is replaced with worst-case first delay.

This observation is important because it ensures our results still apply to
schemes that do lazy traversal of the unreachable space~\citep{lazyrecovery}.
For example, consider a GC guaranteeing that whenever there are unreachable
regions, at least \emph{one} of them is on the free list, but not necessarily
all --- some unreachable regions may be missing from the free list until it
becomes empty, at which case the GC goes searching for another unreachable
element to add.\footnote{\citet{lazyrecovery} provides an example of such a
scheme, except that it uses reference counting so it fails to meet even that
guarantee in the presence of cycles (the basic idea is to delay recursive
decrementing of child reference counts until the parent is actually allocated
again).}
Using our normal notion of delay, any lazy scheme meeting that guarantee has
$d(n) = \infty$ because, unless more space is requested, nodes after the head
might never be added to the free list.
But this does not really capture the goal behind delay, because such a scheme
might still ensure that all allocations can be serviced.
However, this scheme would have $d_1(n) = 1$ constant \emph{first delay},
because it guarantees that \emph{something} is added to the free list
immediately once \emph{anything} first becomes unreachable.

In summary, our results actually imply that there are pathological programs for
which no GC (that can be adapted to implement the GCDS interface) can quickly
detect even when \emph{the very first unreachable region} has become
unreachable.
Hence, rephrasing our results in terms of first delay makes them stronger, and
shows how our lower bounds apply even to schemes using lazy reclamation
techniques.

\subsection{Randomized GCDS}
So far in this paper, we have assumed deterministic algorithms.
While it is not a major focus of our results, we now briefly discuss the
question of whether and to what extent our lower bounds apply also to
randomized algorithms.

First, we consider algorithms that make random decisions during their execution
but guarantee correctness regardless of the random decisions used (i.e., the
randomness can only affect performance).
Our bounds apply without modification if the delay and pause times are defined
for the worst-case random choices.
Surprisingly, they also apply without modification when the delay and pause
times are defined for the \emph{best} possible random choice you could make
with only the information read so far from the persistent store; this is
because the lower bounds we reduce against~\citep{lbdc,bestlbdc} are proved in
the cell probe model~\citep{cpm}, i.e., they only care about the sequence of
reads from and writes to the persistent store.

The question becomes more complicated if the algorithm is only required to be
correct with a certain (high) probability over the random decisions, e.g.,
under some very unlikely choices it is allowed to leave certain regions
uncollected or collect regions that are still reachable.
To the best of our knowledge neither of the bounds we reduce against work in
this setting, and we are not aware of a way to extend our results to this
setting.

\section{Motivating Issue 2: Predictability of Finalizers}
\label{app:MotivateFinalizers}
\begin{figure}[t]
\begin{minted}{Lua}
-- (excerpted implementations away of some functions, globals)
local function finalize(descriptor)
  os.remove(descriptor.path)
  slot_open[descriptor.i] = true end
while true do
  -- phase 1: poll for more work
  for i=1,N_SLOTS do
    if slot_open[i] then
      local fname = "file_test/" .. tostring(i) .. ".txt"
      if path_exists(fname) then
        slot_open[i] = false
        table.insert(worklist,
            setmetatable({i=i,path=fname},{__gc=finalize})) end end end
  -- phase 2: drain the work queue
  while #worklist > 0 do process(table.remove(worklist)) end end
\end{minted}
    \caption{Excerpt from the example server program showing deadlock caused by
    delayed finalization.
    The server shown here polls the filesystem looking for new work.
    Each workitem is associated with a finalizer that deletes the corresponding
    file when that workitem becomes unreachable.
    Files in the directory indicate open workitems, so the client program (not
    shown) treats the files as a lock, i.e., waits for files to be deleted from
    that directory before adding more work for this server.
    Unfortunately, Lua's GC heuristics \emph{never} trigger collection for this
    program, resulting in a deadlock.
    }
    \label{fig:MotivatingFinalizersListing}
\end{figure}

Many programming languages allow the use of finalizers, which are functions
attached to memory regions that get run right before the region is collected.
One program that might seem reasonable to a programmer not expecting the
complexity of modern GCs is excerpted
in~\pref{fig:MotivatingFinalizersListing}.
It is a server that polls for files in a directory and processes each one.
The file descriptors are associated with a finalizer that deletes the underlying
file when the descriptor is collected.
A similar client program (not shown) inserts a file then waits for the file to
be deleted (indicating the server has finished processing it) before adding
another.
In this way, the file acts as a lock, communicating to the client when the
server is ready for more work.

Unfortunately, collection delay means that, even when the file descriptor
becomes unreachable, it may not be collected until much later in the program
execution.
Since unlocking is tied to collection of the descriptor, this can unnecessarily
increase lock contention.

Even worse, many popular languages use heuristics that improve GC performance
in typical cases but can break even the $O(n)$ delay guarantee in the worst
case.
Lua is one such language, and we were surprised to see this cause an immediate
deadlock with this system because it performs too few memory operations to
trigger garbage collection.
In this case, the file descriptors from the first work iteration are never
actually finalized, hence the underlying files are never removed, so the
client is never able to add additional work, triggering deadlock.

This scenario is made worse by the fact that the GC is unpredictable.
When the number of work slots is set to 5 collection gets triggered frequently
enough to avoid deadlock.
But if the number of slots set to either, e.g., 1 or 100, deadlock is
encountered.
Similarly, if the code processing each workitem method performs a lot of memory
operations the GC might be triggered frequently, avoiding the deadlock.
But, if the \texttt{process} method is optimized to perform fewer memory
operations, the deadlock can appear unexpectedly.
In a language like Python that supports both reference counting and infrequent
garbage collection, the finalizers might be promptly called for every version
of the program until the programmer introduces a cyclic data structure,
triggering delayed collection and hence either deadlock or increased lock
contention.

It is relatively easy to warn programmers of these issues, and dissuade them
from attempting to use finalizers as a convenient method to detect when regions
are no longer used by the program.
However, finalizers are a useful language feature, and it is interesting to
ask, as we do in this paper, whether there is any efficient way to make them
significantly more predictable.
Unfortunately, our impossibility result proves that this sort of scenario is
impossible to avoid.

\subsection{Implication of Main Lower Bound for Guaranteed Finalization}
Suppose the language attempts to guarantee for the user that finalizers are
called reliably and promptly, i.e., $d(n) = O(1)$.
Such a guarantee would be convenient, and allow users to rely on finalization
for locks, etc., while avoiding the deadlock described above.

Unfortunately, our lower bound in~\pref{sec:LowerBounds1} implies that there
must exist a program, even a program where nothing ever becomes unreachable,
where this language introduces an $\TOmega(\log^{3/2}{n})$-length pause time
after some program operation.
Hence, no collector guaranteeing prompt finalization is suited for all
real-time settings.

\subsection{Delayed Finalization With Immediate GCs}
We tried running the same motivating example using the immediate GCs described
in~\pref{sec:Eval}.
Both our approach and SynCC guarantee immediate finalization, hence avoiding
the deadlock issue.
They take approximately 1.8 and 2 seconds, respectively, to complete the
benchmark, which is comparable to a manually managed version on the unmodified
Lua interpreter that takes approximately 1.8 seconds.
This again highlights the importance of reducing collection delay in GCs.

\renewcommand{\theTotPages}{28}
\renewcommand{\thepage}{28}

\end{document}